\documentclass[a4,12pt]{article}

\usepackage{tabularx} 
\usepackage{amsmath}  
\usepackage{amssymb}  
\usepackage[colorinlistoftodos]{todonotes} 
\usepackage{commath}
\usepackage{braket} 
\usepackage{float} 
\usepackage{color} 
\usepackage{xcolor} 
\usepackage{graphicx} 
\usepackage[margin=1.2in,letterpaper]{geometry} 
\usepackage{cite} 
\usepackage[normalem]{ulem}
\usepackage{bbm}
\usepackage{textcomp}
\usepackage{wasysym}
\usepackage{dsfont}

\usepackage[mathscr]{euscript}
\usepackage{ntheorem}
\usepackage[final]{hyperref} 
\hypersetup{
	colorlinks=true,       
	linkcolor=black,        
	citecolor=blue,        
	filecolor=magenta,     
	urlcolor=blue
}

\newcommand{\meter}{\mathrm{m}}
\newcommand{\myepsilon}{ {\mathring\varepsilon}}
\newcommand{\nodanger}{}
\newcommand{\mdanger}{\mbox{\dangerous}}

\newcommand{\danger}{\mbox{\red{\dangerous}}}

\newcommand{\redad}[2]{\blue{\ptc{change or addition or rewording on #1} #2}}

\newcommand{\nomathringB}{{B}}

\newcommand{\mathringB}{{\mathring B}}
\newcommand{\mathringA}{{\mathring A}}

\newcommand{\smalle}{{ {\mathfrak{e}}}}
\newcommand{\smallh}{{ {\mathfrak{h}}}}

\newcommand{\noatwo}{ {{\beta_1}}}

\newcommand{\plusatwo}{ {{-2}}}
\newcommand{\plustwoatwo}{ {{-4}}}

\newcommand{\aonesquared}{{{\psi_0^2}}}
\newcommand{\aone}{ {{\psi_0}}}

\newcommand{\BobbyE}{\red{\mycal E}}%
\newcommand{\BobbyB}{\red{\mycal B}}

\newcommand{\eeal}[1]{\label{#1}\end{eqnarray}}

\DeclareFontFamily{OT1}{rsfs}{}
\DeclareFontShape{OT1}{rsfs}{m}{n}{ <-7> rsfs5 <7-10> rsfs7 <10-> rsfs10}{}
\DeclareMathAlphabet{\mycal}{OT1}{rsfs}{m}{n}

\newcommand{\Z}{\mathbbm{Z}}

\newcommand{\blue}[1]{{\color{blue} #1}}

\definecolor{applegreen}{rgb}{0.55, 0.71, 0.0}
\definecolor{armygreen}{rgb}{0.29, 0.33, 0.13}

\newcommand{\green}[1]{{\color{caribbeangreen} #1}}

\definecolor{caribbeangreen}{rgb}{0.0, 0.8, 0.6}

\newcommand{\dire}{ {e}}
\newcommand{\dirb}{ {b}}
\newcommand{\dird}{ {d}}
\newcommand{\dirh}{ {h}}

\newcommand{\blueE}{E}
\newcommand{\blueB}{B}

\newcommand{\Ered}{\red{\mathscr{E}}}
\newcommand{\redE}{\Ered}
\newcommand{\Bred}{\red{\mathscr{B}}}
\newcommand{\redB}{\Bred}

\newcommand{\Dgreen}{\green{\mathds D}}

\newcommand{\Hgreen}{\green{\mathds H}}
\newcommand{\zepsilon}{\blue{\mathring{\epsilon}}}

\newcommand{\griem}{\blue{g}}
\newcommand{\Driem}{\blue{\mathcal D}}

\newcommand{\glorentz}{\red{\mathbf g}}
\newcommand{\Dlorentz}{\red{\nabla}}

\newcommand{\ol}[1]{\overline{#1}}

\newcommand{\red}[1]{{\color{red} #1}}

{\catcode `\@=11 \global\let\AddToReset=\@addtoreset}
\AddToReset{equation}{section}

\newcounter{mnotecount}[section]

\renewcommand{\themnotecount}{\thesection.\arabic{mnotecount}}

\newcommand{\mnotex}[1]
{\protect{\stepcounter{mnotecount}}$^{\mbox{\footnotesize
$
\bullet$\themnotecount}}$ \marginpar{
\raggedright\tiny\em
$\!\!\!\!\!\!\,\bullet$\themnotecount: #1} }

\newcommand{\bel}[1]{\begin{equation}\label{#1}}
\newcommand{\bea}{\begin{eqnarray}}
\newcommand{\bean}{\begin{eqnarray}\nonumber}
\newcommand{\beal}[1]{\begin{eqnarray}\label{#1}}
\newcommand{\eea}{\end{eqnarray}}

\newcommand{\nn}{\nonumber}

\newcommand{\be}{\begin{equation}}
\newcommand{\eeq}{\end{equation}}
\newcommand{\ee}{\end{equation}}
\newcommand{\beqa}{\begin{eqnarray}}
\newcommand{\eeqa}{\end{eqnarray}}
\newcommand{\beqan}{\begin{eqnarray*}}
\newcommand{\eeqan}{\end{eqnarray*}}
\newcommand{\ba}{\begin{array}}
\newcommand{\ea}{\end{array}}

\renewcommand{\danger}{}
\renewcommand{\mdanger}{}

\renewcommand{\redad}[2]{#2}

\begin{document}

\title{Weakly gravitating isotropic waveguides\thanks{Preprint UWThPh-2018-9}
}

\author{
R.\ Beig\thanks{Faculty of Physics, University of Vienna},
P.T. Chru\'sciel$^{\dagger}${}\thanks{Research Platform TURIS, University of Vienna}{},
C.\ Hilweg$^{\dagger\ddagger}$, P. Kornreich\thanks{Professor Emeritus, Syracuse University},
and P.\ Walther$^{\dagger\ddagger}$}

\maketitle

\begin{abstract}
We analyse the effect of post-Newtonian gravitational fields on propagation of light in a cylindrical waveguide in both a straight configuration and  a spool configuration. We derive an equation for the dependence of the wave vector upon the vertical location of the waveguide. It is shown that the gravitational field produces a small shift in the wave vector, which we determine, while the spooling creates additional modes which  could perhaps  be measurable in future accurate experiments.
\end{abstract}

\tableofcontents

\section{Introduction and summary}

In a recent paper \cite{HMMMCW} an experiment was proposed to measure the effect of the gravitational field on single photons in a waveguide. The gravitational phase shift of the photon was calculated there by treating  the photon as a particle following a prescribed trajectory in a post-Newtonian metric. The object of this paper is to derive  the phase shift from first principles, using the Maxwell equations in a dielectric medium in such a metric.

The first main result of our work is
the formula which determines the change $\delta \beta$ of the ``wave vector'' $\beta$ as a function of the change $\delta h$ of height of the waveguide:
\begin{equation}\label{28IX17.1+}
   \delta \beta
   = -  \frac{2 \omega^2 n^2 g }{ c^{ 4} \beta}  \delta h \,.
\end{equation}
cf.\ Equation~\eqref{28IX17.1} below.

Our second main result, derived in Section~\ref{s10II18.1}, is the observation that the periodic gravitational potential effectively experienced by a photon moving in a spool creates additional modes, with wave vector shifted by the inverse of the radius of the spool and with amplitude by about ten orders of magnitude smaller than that of the main mode for a spool with a radius of 30 cm. The effect appears to be out of experimental reach today, but could perhaps be detectable in future accurate experiments.

In retrospect, we find that in our context  \eqref{28IX17.1+} can be obtained   by solving the waveguide problem in Minkowski space-time  and then using the formula,
\begin{equation}\label{16X17.1+a}
 \omega \mapsto   \left(1 - \frac{2\phi}{c^2}\right) \omega
 \,,
\end{equation}
as a replacement for $\omega$ in the Minkowski-space waveguide equations.
Note that this is neither a standard proper-time redefinition for a static observer, nor the standard gravitational redshift formula. In any case, the simple replacement \eqref{16X17.1+a} is  \emph{not} the correct way to proceed  when gradients of $\phi$ become relevant.

Our calculations do not take into account the rotation of the earth. This effect has been calculated in~\cite{HMMMCW} using a particle model, and has been shown to be  dominant in the experiment. A proper treatment of this effect along the lines of the calculations here remains to be done, we will address this in a near future.

Similarly we do not take into account the fact that a non-uniform gravitational field introduces anisotropies in the medium which depend upon the vertical location of the waveguide.
We expect the resulting effect to be of subleading order, but this is not completely clear. We plan to carry-out a careful analysis of this in the future.

Last but not least, we ignore the effects due to the bending of the coils, as they should be independent of the vertical location of the spool.

\section{Isotropic dielectrics in special relativity}
 \label{s29I18.1}

In this section we review the Maxwell equations, in four-dimensional tensor notation, in a charge-free, current-free, non-magnetic, isotropic dielectric medium in Minkowski space-time.

Let
\begin{equation}\label{23VI17.1}
 \eta\equiv \eta_{\alpha \beta} dx^\alpha dx^\beta\equiv  -c^2 dt^2 + dx^2+dy^2+dz^2
\end{equation}
denote the Minkowski metric. Here we use the notation
$$
 (x^\mu)\equiv (x^0,x^i)\equiv (t,x,y,z)
$$
for the coordinates. (In particular, in our conventions, $x^0=t$ and \emph{not} $x^0=ct$, and we are mostly using SI units in this work.) Thus
$$
 \mbox{$(\eta_{ {\alpha\beta}})=\mathrm{diag}(-c^2,1,1,1)$ and $(\eta^{ {\alpha\beta}})=\mathrm{diag}(-c^{-2},1,1,1)$.}
$$

Consider a family of inertial observers in Minkowski spacetime with four-velocity field $u^\mu\partial_\mu = {c^{-1}}\partial_0$.
The Maxwell fields $\vec E $ and $\vec B$ seen by those observers can be encoded in an antisymmetric tensor field $F_{ {\alpha\beta}}$ by setting
\begin{equation}
 \label{8VI17.1}
 F^{0i}=   c^{-2} E^i
 \,,
 \quad
 F_{ij} = \mathring\epsilon_{ijk} B^k
 \,,
\end{equation}
where the indices on $F$ are raised and lowered with $\eta$.
The equations
\begin{equation}
 \label{8VI17.2}
 \partial_{[\alpha}F_{\beta\gamma]} =0
  \quad
  \Longleftrightarrow
  \quad
  d(\underbrace{F_{\alpha \beta} dx^\alpha \wedge dx^ \beta}_{=:F}) = 0
\end{equation}
coincide with the following pair of Maxwell equations
\begin{equation}\label{12VI17.3}
 \nabla \cdot \vec B = 0
 \,,
 \quad
 \partial_t \vec B = - \nabla \wedge \vec E
 \,.
\end{equation}

Let $\zepsilon{}_{ijk}$ and $\zepsilon{}^{ijk}$ denote totally anti-symmetric three-dimensional tensors (without any weights), thus
%
$$
 \zepsilon{}_{123} = 1 = \zepsilon{}^{123}
 \,.
$$
Introducing
\begin{equation}
 \label{8VI17.1+}
 \overline F^{0i}=  {c^{-2}} D^i
 \,,
 \quad
 \overline F_{ij} =  {c^{-2}} \mathring\epsilon_{ijk} H^k
 \,,
\end{equation}
the remaining source-free Maxwell equations in a non-polarizable, homogenous isotropic dielectric, using SI units for the Maxwell fields,
\begin{equation}\label{12VI17.2}
 \nabla \cdot \vec D= 0
 \,,
 \quad
 \partial_t \vec D =  \nabla \wedge \vec H
 \,,
\end{equation}
 can be written as
\begin{equation}
    \nabla^\nu \overline{F}_{\nu\rho} = 0\,,
\end{equation}
where
\begin{equation}
    \overline{F}_{\nu\rho} = a F_{\nu\rho} - 2(a-b) u_{[\nu} F_{\rho]\sigma} u^\sigma
       \,,
       \label{8VI17.4}
\end{equation}
with
\begin{equation}
  \label{8VI17.3}
   b = \epsilon\,,
    \qquad
    a = \frac{1}{\mu c^2  }
    \,.
\end{equation}
Indeed, using $\eta_{ {\alpha\beta}}u^\alpha u^\beta = -1$, \eqref{8VI17.4} is equivalent to the pair of equations $\overline{F}_{\nu\rho} u^\rho = b \,F_{\nu\rho} u^\rho$ and $u_{[\nu} \overline{F}_{\rho\sigma]} = a \,u_{[\nu} F_{\rho\sigma]}$, which is in turn equivalent to the usual relations for a dielectric
\begin{equation}
  \label{8VI17.5}
   \vec D = \epsilon  \vec E  \equiv \epsilon_0 n^2 \vec E
   \,,
   \quad
   \vec B = \mu \vec H  \equiv  \mu_0 \vec H
   \,.
\end{equation}
Note that we can also write
\begin{equation}
 \label{10VI17.1}
    \overline{F}^{\alpha\beta} = \frac 1 {\mu c^2 } \mathring\gamma^{\alpha \rho} \mathring\gamma^{\beta \sigma} F_{\rho\sigma}\,,
\end{equation}
where $\mathring\gamma^{ {\alpha\beta}}$ is  given by
%
\begin{equation}
 \label{10VI17.2}
 \mathring\gamma^{ {\alpha\beta}} = \eta^{ {\alpha\beta}} + ( 1- n^2)u^\alpha u^\beta\,,\,\,\,\mathrm{with}\ n = \sqrt{\epsilon \mu} c
 \,.
\end{equation}
When viewed as a matrix, $\mathring\gamma^{ {\alpha\beta}}$ is the matrix inverse to the \emph{optical metric} $ \mathring\gamma \equiv  \mathring\gamma_{ {\alpha\beta}} dx^\mu dx^\nu$:
\begin{equation}
 \label{10VI17.2-}
  \mathring\gamma_{ {\alpha\beta}} := \eta_{ {\alpha\beta}} + (1- n^{-2})u_\alpha u_\beta
 \,.
\end{equation}

\section{Isotropic dielectrics in curved spacetime}
 \label{s23VII17.2}

We invoke the correspondence principle to write the Maxwell equations for an isotropic dielectric medium in a general space-time, compare~\cite{PerlickBook} and references therein.%
\footnote{As emphasised e.g.\ in~\cite{Hehl2016}, Maxwell equations in media can be written in a covariant way without invoking a metric. In our case the metric arises in the problem at hand through the definition of an isotropic medium.}
The homogeneous Maxwell equations contained in $dF=0$ are already generally covariant and therefore remain unchanged. In the remaining equations  the partial derivatives are replaced by covariant ones. If we denote by $\mathbf{g}$ the space-time metric and by $\nabla$ the associated covariant derivative then,  in the absence of sources,
\begin{equation}
 0=\nabla_\mu \bar{F}^{\mu \nu} \equiv \frac{1}{\sqrt[]{-\det{\red{\mathbf{g}}}}} \frac{\partial (\bar{F}^{\mu \nu} \sqrt[]{-\det{\red{\mathbf{g}}}})}{\partial x^\mu}.
\end{equation}
Equation~\eqref{10VI17.1} maintains its special-relativistic form,
\begin{equation}
 \label{10VI17.1a}
    \bar{F}^{ {\alpha\beta}} = \frac 1 {\mu c^2 } \gamma^{\alpha \rho} \gamma^{\beta \sigma} F_{\rho\sigma}\,,
\end{equation}
with $\gamma^{ {\alpha\beta}}$ defined as
\begin{equation}
 \label{10VI17.2a}
 \gamma^{ {\alpha\beta}} = \red{\mathbf{g}}^{ {\alpha\beta}} + ( 1- n^2)u^\alpha u^\beta\,\,\,\,\mathrm{with}\ n = \sqrt{\epsilon \mu} c
 \,.
\end{equation}
Here the vector field $u$ is taken to be the velocity four-vector defined by the motion of the dielectric medium in space-time. It depends of course upon $\red{\mathbf{g}}$ via its normalization $\red{\mathbf{g}}(u,u)=u^\alpha u_\alpha = -1$. Equivalently, assuming that $\dot x\equiv \dot x^\alpha \partial_\alpha$ is future-oriented,
\begin{equation}
u ^\gamma := \frac{\dot x^\gamma}{\sqrt{\abs{\red{\mathbf{g}}(\dot x, \dot x)}}}
 \equiv
 \frac{\dot x^\gamma}{\sqrt{\abs{\red{\mathbf{g}}_{\alpha\beta}\dot x^\alpha \dot x^\beta}}}
 \,.
\end{equation}

\subsection{$3+1$ decomposition}
 \label{ss16VII17.3}

In this section we rewrite the Maxwell equations in a $3+1$--decomposed form. For future reference we do this in a whole generality, for any metric  and in any coordinate system. Note that for the purpose of our work here a static metric in an adapted time slicing would have sufficed. A geometric decomposition of the equations in this last setting can be found in Appendix~\ref{A11II18.1}.

Let   us set
\bel{16VII17.11}
 \dire_i := F_{i0}
 \,,
 \quad
 \dird^i := \ol F^{0i}
 \,,
 \quad
 \dirb^k
  := \frac 12 \zepsilon{}{}^{ijk} F_{ij}
 \,,
 \quad
 \dirh_k
  := \frac 12 \zepsilon{}{}_{ijk} \ol F^{ij}
   \,.
\ee
In Minkowski space-time and in a manifestly Minkowskian coordinate system we have, using SI units,
\bel{16VII17.11+}
 \dire_i := E_i
 \,,
 \quad
 \dird^i := c^{-2} D^i
 \,,
 \quad
 \dirb^k
  := B^k
 \,,
 \quad
 \dirh_k
  := c^{-2} H^k
   \,.
\ee

In the notation of \eqref{16VII17.11}, the equations
\bel{16VII17.12}
 \partial_{[\alpha} F_{\beta\gamma]} =0
\ee
are equivalent to
%
\bel{16VII17.13}
 \partial_{0} \dirb^k = - \zepsilon{}{}^{ijk} \partial_i \dire_j
 \,,
 \quad
  \partial_k \dirb^k = 0
 \,.
\ee
Similarly, the equations
\bel{16VII17.21}
 \Dlorentz_\mu \ol F^{\mu\nu} = 0
 \,,
\ee
%
decompose as
%
\bel{16VII17.23}
 \partial_{0}(\sqrt{|\det \glorentz|}  \dird^k )=  \partial_i(\sqrt{|\det \glorentz|}  \zepsilon{}{}^{ijk}\dirh_j
 )
 \,,
 \quad
  \partial_k (\sqrt{|\det \glorentz|}  \dird^k ) = 0
 \,.
\ee

Let $\griem= \griem_{ij}dx^i dx^j$ denote the metric induced by the space-time metric on the level sets of the function $x^0$, and let $\Driem$ denote the associated covariant derivative.
In order to make \eqref{16VII17.13} manifestly covariant with respect to space-coordinate changes we introduce
\begin{equation}\label{16VII17.14abc}
  \epsilon_{ijk} = \sqrt{\det \griem } \zepsilon_{ijk}
  \,,
  \quad
  \epsilon^{ijk} = \frac{1}{\sqrt{\det \griem }} \zepsilon^{ijk}
  \,,
  \quad
  \Bred^k = \frac 12 \epsilon^{ijk} F_{ij} = \frac{1}{\sqrt{\det \griem }} \dirb^k
   \,.
\end{equation}
This allows us to rewrite \eqref{16VII17.13} as
%
\bel{16VII17.14abd}
 \fbox{$
  \partial_{0} \Bred^k = - \epsilon{}{}^{ijk} \Driem_i \dire_j
 -\frac 12 \frac{\partial_0 (\det \griem) }{\det \griem} \Bred^k
 \,,
 \quad
  \Driem_j \Bred^j = 0
 \,.
 $}
\ee
Let $c$ be a constant related to the choice of units, which has its usual meaning in a post-Newtonian context (thus, in Minkowski space-time and in SI units, equals the speed of light).
Setting
%
\begin{equation}\label{16VII17.32-}
  \Dgreen^k:= \frac{c \dird^k}{\sqrt{|\glorentz^{00}|}}
  \,,
   \quad
  \Hgreen_k:= \frac{{\sqrt{\det \griem}}}{\sqrt{|\glorentz^{00}|}} c \dirh_k \equiv \sqrt{|\det \glorentz|}
    c \dirh_k
  \,,
\end{equation}
where we used
\begin{equation}\label{16VII17.31}
  \det \glorentz = \frac {\det \griem}{\glorentz^{00}}
  \,,
\end{equation}
the space-covariant version of \eqref{16VII17.23} reads
%
\bel{16VII17.32}
 \fbox{$
 \partial_{0}  \Dgreen^k  = \epsilon{}{}^{ijk} \Driem_i\Hgreen_j
 - \frac 12 \frac{\partial_0 (\det \griem) }{\det \griem} \Dgreen^k
 \,,
 \quad
  \Driem_k  \Dgreen^k  = 0
 \,.
 $}
\ee

So far the equations were completely general, and holding in any coordinate system. In what follows we will now assume that the four-velocity vector field   $u^\mu$ of the medium is hypersurface-orthogonal. We can then use a coordinate system in which
$u_\alpha dx^\alpha$ is proportional to $dt$ and  $u^\beta \partial_\beta$ is proportional to $\partial_t$.%
\footnote{One could moreover require $u^\alpha\partial_\alpha=\partial_t$, but this is not convenient when, e.g., a post-Newtonian metric is used.}
Assuming \eqref{10VI17.1}-\eqref{10VI17.2} we then have
%
\begin{equation}\label{16VII17.41}
  \dird^k =  \epsilon |\glorentz^{00}| g^{k\ell} \dire_\ell
   \,,
   \quad
  \Dgreen^k= c \epsilon  {\sqrt{|\glorentz^{00}|}} g^{k\ell} \dire_\ell
  \,,
   \quad
   \Bred^k = \mu c \sqrt{|\glorentz^{00}| } g^{k\ell} \Hgreen_\ell
   \,.
\end{equation}
%
Setting
\begin{equation}\label{16VII17.51}
  \redE^k :=  c\sqrt{|\glorentz^{00}|} g^{k\ell}\dire_\ell
  \,,
\end{equation}
%
one is led to
%
\begin{eqnarray}
 \label{16VII17.14a}
 &
  \displaystyle
 \partial_{t} \Bred^k =
   - \epsilon{}{}^{ijk} \Driem_i
     \left(\frac{\redE_j}{ c\sqrt{|\glorentz^{00}|} }
      \right)
-\frac 12 \frac{\partial_t (\det \griem) }{\det \griem} \Bred^k
 \,,
 \quad
  \Driem_j \Bred^j = 0
 \,,
 &
\\
 &
  \displaystyle
 \partial_{t} \left( \epsilon \redE^k\right) = \epsilon{}{}^{ijk}
 \Driem_i\left( \frac{\Bred_j }{\mu c \sqrt{|\glorentz^{00}|}}\right)
 - \frac \epsilon {2 } \frac{\partial_t (\det \griem) }{\det \griem} \Ered^k
 \,,
 \quad
  \Driem_k \left(  \epsilon \redE^k \right)   = 0
 \,;
  \phantom{xxx}
   &
 \label{16VII17.14b}
\end{eqnarray}
recall that all indices are raised and lowered with respect to the space-metric $g_{ij}dx^i dx^j$.

Note that both $\epsilon$ and $\mu$ are allowed to depend upon coordinates as well as all other fields.

\subsection{Linearised gravity}

In the weak-field approximation the metric tensor $\red{\mathbf{g}}_{\mu \nu}$ is written as
\begin{equation} \label{linearised}
\red{\mathbf{g}}_{\mu \nu} = \eta_{\mu \nu} + h_{\mu \nu},
\end{equation}
where $\eta_{\mu \nu}$ is the Minkowski metric and $h_{\mu \nu}$ is a \textit{small}  deviation. By definition, there exists a small number $\myepsilon>0$ (not to be confused with the vacuum permittivity $\epsilon_0$) such that
\begin{equation} \label{linearisedmetricepsilon}
 |h| + |\partial h| + |\partial \partial h| \le \myepsilon \ll 1
 \,.
\end{equation}
In the calculations that follow we will ignore all terms  involving products of $h$ and its derivatives, i.e. any terms of order 2 or higher in $\myepsilon$.  In such an approximation it holds that
\begin{equation}
\red{\mathbf{g}}^{\mu \nu} = \eta^{\mu \nu} - h^{\mu \nu} + O (\myepsilon ^2)
 \,,
\end{equation}
while the determinant of the  metric $\red{\mathbf{g}}_{\mu \nu}$ can be written as
\begin{equation} \label{morecompactdet}
 \det{\red{\mathbf{g}}} = - c^2 (1+ h^\alpha{}_\alpha + O(\myepsilon^2))
 \quad
 \Longrightarrow
  \quad
 \sqrt{|\det{\red{\mathbf{g}}}|} =  c (1+  \frac 12 h^\alpha{}_\alpha + O(\myepsilon^2))
 \,.
\end{equation}

\subsection{Post-Newtonian approximation}
 \label{ss16VII17.1}

We consider \eqref{16VII17.14a}-\eqref{16VII17.14b} in the weak-field post-Newtonian metric $\red{\mathbf{g}}_{\mu\nu} = \eta_{\mu\nu} + h_{\mu\nu}$, thus
%
\begin{equation}
 \label{10VI17.12p}
 h_{00} =   - 2 \phi
 \,,
 \quad
 h_{ij} =  - \frac{2 \phi}{  c^2} \eta_{ij}
 \,,
 \quad
 h_{0i} =   0
 \,,
 \quad
  \frac{1 }{\sqrt{|\glorentz^{00}|}} =  c \left(
    1+ \frac{  \phi}{  c^2}
     \right)
 \,,
 \quad
 \epsilon^{ijk} = (1+ \frac{3 \phi}{  c^2}) \zepsilon^{ijk}
 \,,
\end{equation}
where $\phi$ is the Newtonian potential; in our conventions,
$$
 \phi=- MG/r
$$
in spherical symmetry.  This is equivalent to using a Schwarzschild metric, and calculating to leading order in the inverse-distance. In this section we assume that the dielectric is at rest in the coordinates above. Using $\Ered_i = g_{ij} \Ered^j =(1-2\phi c^{-2}) \Ered^i$, etc.,
\redad{28VI18}{
and using the summation convention on repeated pairs of indices even if they are in identical positions,
}
 in the regions where $\mu$ and $\epsilon$ are constant from \eqref{16VII17.14a}-\eqref{16VII17.14b} we find
%
\begin{eqnarray}
 \label{16VII17.14ap}
 &
  \displaystyle
 \partial_{t} \Bred^k = -  (1+ \frac{3 \phi}{  c^2}) \zepsilon^{ijk}
 \partial_i
  \left(( 1- \frac{  \phi}{  c^2}  ) \Ered^j  \right)
 \,,
 \quad
  \partial_j\left( (1-  \frac{3 \phi}{  c^2}) \Bred^j
   \right)
   = 0
 \,,
 &
\\
 &
  \displaystyle
  {\epsilon \mu}
 \partial_{t} \redE^k  =  (1+ \frac{3 \phi}{  c^2}) \zepsilon^{ijk}
 \partial_i\left(( 1- \frac{   \phi}{  c^2}  ) \Bred^j  \right)
 \,,
 \quad
  \partial_j\left( (1-  \frac{3 \phi}{  c^2}) \redE^j
   \right)
   = 0
 \,.
   &
 \label{16VII17.14bp}
\end{eqnarray}
To leading order we obtain
\begin{eqnarray}
 \label{16VII17.14ap1}
 &
  \displaystyle
 \left(
  1- \frac{2 \phi}{  c^2}
  \right)
   \partial_{t} \Bred^k = -   \zepsilon^{ijk} \partial_i  \Ered^j
  +  \zepsilon^{ijk}\frac{   \partial_i \phi}{  c^2}   \Ered^j
 \,,
 \quad
  \partial_j\left( (1-  \frac{3 \phi}{  c^2}) \Bred^j
   \right)
   = 0
 \,,
 &
\\
 &
  \displaystyle
 \left(
  1- \frac{2 \phi}{  c^2}
  \right)
    {\epsilon \mu}
 \partial_{t} \redE^k  = \zepsilon^{ijk}
 \partial_i  \Bred^j
 -   \zepsilon^{ijk}
  \frac{ \partial_i \phi}{  c^2}   \Bred^j
 \,,
 \quad
  \partial_j\left( (1-  \frac{3 \phi}{  c^2}) \redE^j
   \right)
   = 0
 \,.
   &
 \label{16VII17.14bp1}
\end{eqnarray}
Assuming that $\phi$ is constant throughout the dielectric, we find the usual equations for a dielectric with the original $n$ but with a redefinition of time,
$$
 \fbox{$\displaystyle  t\mapsto  \left(
  1 + \frac{2 \phi}{  c^2}
  \right) t
  $
  }
 \,.
$$
Note that this differs by a factor of two from the usual \emph{time-dilation} arising from a redefinition of coordinate time $t$ to proper time of static observers, which corresponds to (in leading order in $\phi/c^2$)
$$
 t\mapsto  \left(
  1 + \frac{  \phi}{  c^2}
  \right) t
 \,.
$$
For practical purposes it is advantageous to introduce new fields
\begin{equation}\label{27VII17.3-}
 \blueE^j = (1-  \frac{3 \phi}{  c^2}) \redE^j
 \,,
 \quad
 \blueB^j  =(1-  \frac{3 \phi}{  c^2}) \Bred^j
 \,,
\end{equation}
which have vanishing Euclidean divergence.

(Note that the notation is in line with that of Section~\ref{s29I18.1} when $\phi \equiv 0$.)

In terms of these we find
%
%
\begin{eqnarray}
 \label{27VII17.3}
 &
 \fbox{$
 \left(
  1- \frac{2 \phi}{  c^2}
  \right)
   \partial_{t} \blueB^k = -   \zepsilon^{ijk} \partial_i  \blueE^j
   $}
   -
  2  \zepsilon^{ijk}\frac{   \partial_i \phi}{  c^2}    \blueE^j
 \,,
 \quad
  \partial_j \blueB^j
   = 0
 \,,
 &
\\
 &
 \fbox{$
 \left(
  1- \frac{2 \phi}{  c^2}
  \right)
 \epsilon \mu
 \partial_{t}  \blueE^k  = \zepsilon^{ijk}
 \partial_i  \blueB^j
 $}
 +2
    \zepsilon^{ijk}
  \frac{ \partial_i \phi}{  c^2}   \blueB^j
 \,,
 \quad
  \partial_j\blueE^j
   = 0
 \,.
   &
 \label{27VII17.4}
\end{eqnarray}
%
The boxed part of the equation is the relevant one when $\phi$ is assumed to be constant throughout the dielectric.

We emphasise that above we use $\blueE^j$ and $\blueB^j$ as the basic fields because they have vanishing divergence. 

Differentiating with respect to $t$ and neglecting terms involving more than one factor of $\phi$ one finds, in Euclidean coordinates on $\mathbbm{R}^3$,
%
\begin{eqnarray}
 \label{27VII17.3asdf}
 &
 \fbox{$ \left(
  1- \frac{2 \phi}{  c^2}
  \right)^2
  {\epsilon \mu}
   \partial_{t}^2  \blueB^k =   \mathring \Delta \blueB^k
   $}
   + \frac 2 {c^2} (\blueB^k\mathring \Delta \phi - \blueB^i \partial_k \partial_i \phi)
   + \frac 1 {c^2} \partial^i \phi(\red{6}\partial_i  \blueB^k - \red{4} \partial_k \blueB^i)
 \,,
  \qquad
 &
\\
 &
 \fbox{$
  \left(
  1- \frac{2 \phi}{  c^2}
  \right)^2
  {\epsilon \mu}
   \partial_{t}^2  \blueE^k =   \mathring \Delta \blueE^k
   $}
   + \frac 2 {c^2} (\blueE^k\mathring \Delta \phi - \blueE^i \partial_k \partial_i \phi)
   + \frac 1 {c^2}    \partial^i \phi (\red{6}\partial_i  \blueE^k - \red{4} \partial_k \blueE^i)
 \,,
  \qquad
   &
 \label{30VII17.asdf1}
\end{eqnarray}
where $\mathring \Delta$ is the flat-space Laplace operator $\partial_x^2 + \partial_y^2 + \partial_z^2$.

Neglecting terms involving one or more derivatives  of $\phi$, we obtain the usual second-order propagation equation in the medium  with $n$ replaced by an ``effective refractive index''
\begin{equation}\label{16VII17.61}
 n _{\mathrm{eff}}:=  (1- \frac{2  \phi}{  c^2}) \sqrt{\epsilon \mu} c  > \sqrt{\epsilon\mu} c
    \,.
\end{equation}
However, this is \emph{not} the right interpretation for \eqref{27VII17.3}-\eqref{27VII17.4}.

\subsection{Orders of magnitude}
 \label{ss3VIII17.1}

On the surface of the earth we have: mass ${M_{\earth} \mdanger \approx}  5.97  \times 10^{24} \mathrm{kg}$, radius ${R_{\earth} \mdanger \approx }  6.37 \times 10^6 \mathrm m$, Newton's constant $ G\approx 6.67  \times 10^{-11}{\mathrm {\ m}^{3}\ \mathrm{kg}^{-1}\ \mathrm{s}^{-2}}$, speed of light $c\approx  3 \times 10^ 8 \mathrm m/\mathrm s$, giving
\begin{equation}\label{16VII17.62}
  \frac{2 \phi}  {c^2} \approx  - 1.39 \times 10^{-9}
  \,.
\end{equation}
For the purpose of comparing waveguides at different heights this is by itself not relevant (compare the beginning of Section~\ref{ss26IX17.2} below), what matters is only the difference of the potential between the waveguides, and hence the gradient of $\phi$.
Denoting by
$ g_{\earth} \approx9.81 \mathrm m / \mathrm s^2$  the gravitational acceleration at the surface of the earth, we find
\begin{equation}\label{16VII17.63}
  \frac{|\nabla \phi|} {c^2} =  g_{\earth}  c^{-2} \approx 1.09 \times 10^{-16} \mathrm {\mathrm{m}^{-1}}
  \,.
\end{equation}

When used together with \eqref{16VII17.61},
one obtains the relative change of effective refractive index when a second waveguide is located one meter above the first one.  Note that this relative change can be increased by two orders of magnitude by placing the upper waveguide at the top of the ZARM tower in Bremen, with the lower arm at the bottom.

Note that, in view of \eqref{16VII17.62}, corrections in the equations of relative order beyond $10^{-18}$ might require taking into account second-order post-Newtonian corrections.

In order to estimate the relative effect of the gradient terms in \eqref{16VII17.14ap1}-\eqref{16VII17.14bp1}, the gradient \eqref{16VII17.63}
should be compared with $ |\nabla \vec B|/|\vec B|$. Now, the diameter $d$ of a typical dielectric is
of the order of $d\sim 10^{-5}$m. Estimating $ |\nabla \vec B|/|\vec B|$ by $d^{-1}$, this gives a contribution of the $\partial_i \phi$ terms in the equations smaller by a factor of $10^{-21}$ than the dominating terms in the equations.

Recall that the standard optical fibers are silica fibers, which have very high attenuation for high frequencies. There
are two low loss windows around
\bel{23IV14.5}
  \mbox{$1.94 \times 10^{14} s^{-1}$ and $2.29\times 10^{14} s^{-1}$. }
\ee
(One should explore the possibility of using photonic crystal fibers, but those do not seem to be available at this stage at the required lengths, and are likely to be much more expensive than the silica fibers which are off-the-shelf.)

Now, one can alternatively  estimate $ |\nabla \vec B|/|\vec B|$ by the inverse wave-length $\lambda^{-1}$ of the light in a typical dielectric, which in view of \eqref{23IV14.5} is of order
\begin{equation}\label{11IX17.1}
 \lambda \sim {10^{-6} \mathrm{m}}
 \,.
\end{equation}
This leads to a contribution of the gradient terms of $\phi$ smaller by a factor $10^{-22}$ than the dominant ones, which is of the same order as the corrections in the equations arising from a ``very strong'' gravitational wave of astrophysical origin.

Consider, next, the curvature of a coil. If the spool has a diameter of the order of $10^{-1} m$, a naive guess is that the relative curvature effects might be of the order wavelength/diameter, and hence $10^{-5}$. This is much bigger than the gravitational effect, and therefore a careful analysis of the contribution would be desirable. It is known that curvature of the fibers leads to losses, but an exhaustive analysis at the desired level of accuracy would be useful. One should nevertheless note that the effect should be the same in each arm of an interferometer, and therefore the fibre-curvature effects should not affect the outcome of the experiment proposed in \cite{HMMMCW}.

\section{Cylindrical waveguide in a constant  post-Newtonian potential}
 \label{s26IX17.1}

\subsection{Assumptions and setup}
 \label{ss26IX17.1}

We want to find the modes of an optical fiber in the presence of a weak gravitational field. We will make the following assumptions about the dielectric:

\begin{itemize}
\item  a linear, isotropic, non-magnetic dielectric, without free charges or currents,
and without surface currents;
\item
the dielectric constant $\epsilon$
does not depend upon the vertical location  of the waveguide in the lab;
\item  a perfect step-index fiber, which is homogeneous in the core and in the cladding. The only discontinuity is at the core-cladding interface;
\item we will use cylindrical coordinates, where the $z$-axis is aligned along the symmetry axis of the optical fiber;
\item the medium is lossless.
\end{itemize}

We consider a dielectric with the following constitutive equations:
\
\begin{align}
 \vec{D}(\vec{r},t)&=\epsilon_0n(r)^2\vec{E}(\vec{r},t) \label{equ:consD}
  \,,
\\
 \vec{B}(\vec{r},t)&=\mu_0\vec{H}(\vec{r},t)
  \,.
 \label{equ:consB}
\end{align}
where $n(r)$ denotes the refractive index as a function of the radial coordinate.

\subsection{Post-Newtonian Maxwell equations}
 \label{ss26IX17.2}

We start with a gravitational potential that is constant throughout the fiber, as this will be needed for analysis later in any case. The calculations in this section are a straightfoward adaptation of the textbook ones to the problem at hand, cf.\ e.g.\ \cite{Keiser}.

We will work in SI units where the modified Maxwell equations (ME) for the material as specified above, and for a constant potential, are given by the following, reference-normalised version, of \eqref{27VII17.3}-\eqref{27VII17.4}:
\begin{align}
\nabla \times \vec{E}(\vec{r},t)&=-\psi \partial_t \vec{B}(\vec{r},t) \label{equ:Fara}
 \,,
\\
  \nabla \times \vec{H}(\vec{r},t)&=\psi \partial_t \vec{D}(\vec{r},t) \label{equ:Max}
 \,,
\\
\nabla \cdot \vec{D}(\vec{r},t)&=0 \label{equ:divD}
 \,,
\\
  \nabla \cdot \vec{B}(\vec{r},t)&=0
  \,.
   \label{equ:divB}
\end{align}
Here we defined
\begin{equation}\label{19X17.1}
 \psi:=\frac{1-\frac{2\phi}{c^2}}{1-\frac{2\mathring \phi}{c^2}}
 \,,
\end{equation}
with the gravitational potential, for $R\ge R_{\earth}$,
$$
 \phi=-\frac{G M_{\earth} }{R}
 \,,
$$
where $R$ is the distance to the center of the Earth.
Here $\mathring \phi$ is a normalising constant arising from our choice of coordinates so that the metric takes the manifestly Minkowskian form at the location of a reference waveguide which is used to calibrate the constants at hand.

Some comments on \eqref{19X17.1}  are in order. A direct application of the equations \eqref{27VII17.3}-\eqref{27VII17.4}, derived in the post-Newtonian approximation, where we neglect now the gradients of the gravitational potential, would give $ \psi= 1-\frac{2\phi}{c^2}$. Now, if we denote by $\mathring x$ the position of the center of the lab, and introduce local inertial coordinates associated with the center of the lab, then to compare with special relativity it is useful to rescale the coordinates in the usual post-Newtonian line element,
$$
    ds^2 = -\big (1+\frac{2\phi}{c^2}\big) c^2 dt^2 + \big (1-\frac{2\phi}{c^2}\big) \big(dx^2+ dy^2 + dz^2\big)
    \,,
$$
so that the metric takes the Minkowskian values $\mathrm{diag}(-c^2,1,1,1)$ at the center of the lab:
\begin{equation}\label{15XI17.1}
 t\mapsto (1+\frac{2\mathring \phi}{c^2}\big)^{-1/2}  t
  \,,
  \quad
  x^i \mapsto (1-\frac{2\mathring \phi}{c^2}\big)^{-1/2}  x^i
  \,,
\end{equation}
where $\mathring \phi = \phi(\mathring x)$. Applying this rescaling to \eqref{27VII17.3}-\eqref{27VII17.4} one obtains formula \eqref{19X17.1} for $\psi$.

%

Now, in a first approximation we will be interested  in the effect resulting from the dependence of   $\psi$ upon the height $h$, measured with respect to the reference waveguide used to normalise $\psi$, at which the dielectric has been placed, in which case one can take
\begin{equation}\label{15I18.1}
 \psi= \psi(h)=1+2  g_{\earth} c^{-2} h
 \,,
\end{equation}
where $ g_{\earth} $
denotes again the gravitational acceleration at the surface of the earth.

An alternative way of viewing \eqref{19X17.1}
is to rewrite this equation as
\begin{equation}\label{19X17.1+}
 \psi=\frac{1-\frac{2\phi}{c^2}}{1-\frac{2\mathring \phi}{c^2}}
  = 1-\frac{2(\phi-\mathring \phi)}{c^2}
  + O (\frac{\mathring \phi(\phi-\mathring \phi)}{c^4})
 \,,
\end{equation}
which makes it clear in which sense one  can,
and we will,
write $\phi$ for $\phi-\mathring \phi$ in the equations that follow.

\subsection{The $z$-components of the fields}
 \label{ss3VII18.1}

Each Cartesian component $E^i$ and $H^i$ of $\vec E$ and $\vec H$ satisfies the wave equation \eqref{30VII17.asdf1}. We will seek solutions of the form
\begin{equation}\label{29I18.2}
  \vec E(t,\vec x) = e^{i\omega t} \vec E(\vec x)
  \,,
  \qquad
  \vec H(t,\vec x) = e^{i\omega t} \vec H(\vec x)
  \,,
\end{equation}
with
\begin{equation}\label{1II18.7+}
 \omega\equiv kc = \frac{k}{\sqrt{\epsilon_0\mu_0}}
 \,,
\ee
for a constant $k$.
(We beg the indulgence of the reader with our 
use of the same symbol $\vec E$ for fields depending upon different independent variables; similarly for $\vec H$.)

For the component $E^z$  we use the ansatz
\begin{equation}\label{29I18.1}
   E^z (t,\vec x) = e^{i(\omega t- \beta z- m\theta )} R(r)
  \,,
\end{equation}
and note that a similar ansatz 
will be used for $E^x$ and $E^y$.
Using cylindrical coordinates,  \eqref{30VII17.asdf1} becomes the following equation for $R(r)$:
\begin{align}
&\frac{d^2R}{dr^2}+\frac{1}{r}\frac{dR}{dr}+\left(\psi^2n_1^2k^2-\beta^2-\frac{m^2}{r^2}\right)R=0
 \,, \qquad \text{core,}\\
&\frac{d^2R}{dr^2}+\frac{1}{r}\frac{dR}{dr}+\left(\psi^2n_2^2k^2-\beta^2-\frac{m^2}{r^2}\right)R=0 \,, \qquad \text{cladding.}
\end{align}
For guided modes we must have that the propagation constant in the z-direction is smaller than the wave number  in the core, $\beta<\psi n_1k$, and greater than the wave number  in the cladding, $\beta>\psi n_2 k$.

 Let $a$ denote the coordinate-radius of the core.
which is related to its physical radius $a_p$ as
\begin{equation}\label{a19VI18.1}
  a_p = \sqrt \psi  a
  \,.
\end{equation}
 It is customary to define two new quantities,
\begin{equation}
U^2:=a^2(\psi^2n_1^2k^2-\beta^2) \label{equ:U}
 \,,
\end{equation}
\begin{equation}
W^2:=a^2(\beta^2-\psi^2n_2^2k^2)
 \,,
   \label{equ:W}
\end{equation}
which leads to the following form of the equations in the core and the cladding:
\begin{align}
&\frac{d^2R}{dr^2}+\frac{1}{r}\frac{dR}{dr}+\left(\frac{U^2}{a^2}-\frac{m^2}{r^2}\right)R=0
 \,,
  \qquad \text{core,}   \label{equ:Core}\\
&\frac{d^2R}{dr^2}+\frac{1}{r}\frac{dR}{dr}-\left(\frac{W^2}{a^2}+\frac{m^2}{r^2}\right)R=0
 \,, \qquad \text{cladding.} \label{equ:Cladding}
\end{align}
The solutions of \eqref{equ:Core} are linear combinations of Bessel functions of the first ($J_m(Ur/a)$) and second ($Y_m(Ur/a)$) kind, where we need to reject the $Y_m(Ur/a)$ solution due to their divergence at $r=0$.
The solutions of \eqref{equ:Cladding} can similarly be written as linear combinations of modified Bessel functions of the first ($I_m(Wr/a)$) and second kind ($K_m(Wr/a)$), where we must reject the $I_m(Wr/a)$'s because they diverge  as $r\rightarrow \infty$, which would lead to infinite energy.
The continuous solutions take the form
\
\begin{equation}
\label{equ:SolCoreE}
E^z(r,\theta)=
\begin{cases}
    A\frac{J_m(Ur/a)}{J_m(U)}e^{im\theta} \,,             & \text{core,}\\
    A\frac{K_m(Wr/a)}{K_m(W)}e^{im\theta} \,,             & \text{cladding.}
\end{cases}
\end{equation}
where we have assumed that the denominators do not vanish:
\begin{equation}\label{29I18.3}
  J_m(U) , K_m(W) \ne 0
  \,.
\end{equation}

Using the same ansatz for the $z$-component of the magnetic field we can write the solutions as
\
\begin{equation}
\label{equ:SolCoreH}
H^z(r,\theta)=
\begin{cases}
    B\frac{J_m(Ur/a)}{J_m(U)}e^{im\theta}  \,,            & \text{core,}\\
    B\frac{K_m(Wr/a)}{K_m(W)}e^{im\theta}    \,,          & \text{cladding.}
\end{cases}
\end{equation}

\subsection{The remaining components of the fields}
 \label{ss1II18.1}

We use the post-Newtonian Maxwell equations with constant gravitational potential,
\begin{eqnarray}
 \label{27VII17.3asdf+}
 &
 \psi
    \mu_0 \partial_{t} H ^k = -   \zepsilon^{ijk} \partial_i   E^j
 \,,
\qquad
 \psi
 n^2  \epsilon_0
 \partial_{t}  E^k  = \zepsilon^{ijk}
 \partial_i  H^j
 \,,
   &
\end{eqnarray}
to determine the remaining Cartesian components of $\vec E$ and $\vec B$. The ansatz \eqref{29I18.2} leads to the following equations for the time-independent part of the fields: 
\begin{align}
 \label{28IX17.1asdf}
H^x= \frac{i}{\psi k}\sqrt{\frac{\epsilon_0}{\mu_0}}\left(  \partial_{y}E^z-\partial_z E^{y}\right)
 \,,
  &
\quad
 H^y= \frac{i}{\psi k}\sqrt{\frac{\epsilon_0}{\mu_0}}\left(  \partial_{z}E^x-\partial_x E^{z}\right)
 \,,
\\
E^x= - \frac{i}{\psi k n^2 }\sqrt{\frac{\mu_0}{\epsilon_0}}
 \left(  \partial_{y}H^z-\partial_z H^{y}\right)
 \,,
  &
 \quad
 E^y
  =
    - \frac{i}{\psi k n^2 }\sqrt{\frac{\mu_0}{\epsilon_0}}\left(  \partial_{z}H^x-\partial_x H^{z}\right)
\,.
 \label{28IX17.3asdf}
\end{align}
Inserting \eqref{28IX17.3asdf} into \eqref{28IX17.1asdf} to obtain equations for $H^x$ and $H^y$, and proceeding similarly for the electric field, one finds the following ODEs in $z$:
\begin{align}
 \label{29I18.4}
 H^x
  +
   \frac{1}{\psi k} \partial_z\left(\frac{\partial_z H^x} { \psi k n^2}
    \right)
     = &
     \frac{i}{\psi k}\sqrt{\frac{\epsilon_0}{\mu_0}}  \partial_{y}E^z
  + \frac 1 {\psi k}  \partial_z\left(\frac{\partial_x H^z} { \psi k n^2}
    \right)
 \,,
\\
 H^y
  +
   \frac{1}{\psi k} \partial_z\left(\frac{\partial_z H^y} { \psi k n^2}
    \right)
     = &
     - \frac{i}{\psi k}\sqrt{\frac{\epsilon_0}{\mu_0}}  \partial_{x}E^z
  + \frac 1 {\psi k}  \partial_z\left(\frac{\partial_y H^z} { \psi k n^2}
    \right)
 \,,
\\
E^x
  +
   \frac{1}{\psi k n^2} \partial_z\left(\frac{\partial_z E^x} { \psi k}
    \right)
     = &
       - \frac{i}{\psi k n^2 }\sqrt{\frac{\mu_0}{\epsilon_0}}
 \partial_{y}H^z
  +
   \frac 1 {\psi k n^2} \partial_z \left( \frac{\partial_x E^z}{\psi k}
    \right)
 \,,
\\
 \quad
 E^y
  +
   \frac{1}{\psi k n^2} \partial_z\left(\frac{\partial_z E^y} { \psi k}
    \right)
     =
     &
        \frac{i}{\psi k n^2 }\sqrt{\frac{\mu_0}{\epsilon_0}}
 \partial_{x}H^z
  +
   \frac 1 {\psi k n^2} \partial_z \left( \frac{\partial_y E^z}{\psi k}
    \right)
\,.
 \label{29I18.5}
\end{align}
As particular  solutions of the inhomogeneous ODEs we choose
\begin{align}
 \label{29I18.6}
 H^x
     = &
     \frac{i}{\psi^2 k^2n^2 - \beta^2}
      \left(
        \psi k n^2  \sqrt{\frac{\epsilon_0}{\mu_0}}  \partial_{y}E^z
            - \beta  \partial_x H^z
    \right)
 \,,
\\
 H^y
     = &
    - \frac{i}{\psi^2 k^2n^2 - \beta^2}
      \left(
       {\psi k n^2}\sqrt{\frac{\epsilon_0}{\mu_0}}  \partial_{x}E^z
  +  \beta \partial_y H^z
    \right)
 \,,
\\
E^x
     = &
    - \frac{i}{\psi^2 k^2n^2 - \beta^2}
      \left(
       {\psi k n^2 }\sqrt{\frac{\mu_0}{\epsilon_0}}
 \partial_{y}H^z
  +
   \beta \partial_x E^z
    \right)
 \,,
\\
 \quad
 E^y
     = &
     \frac{i}{\psi^2 k^2n^2 - \beta^2}
      \left(
       \psi k n^2 \sqrt{\frac{\mu_0}{\epsilon_0}}
 \partial_{x}H^z
  - \beta \partial_y E^z
    \right)
\,.
 \label{29I18.7}
\end{align}
%

\subsection{Determining $\beta$}
 \label{ss1II18.2}

 As is well known (cf., e.g.,~\cite{Keiser}), the requirement of continuity of
relevant
fields imposes a condition on $\beta$ and leads to a relation between the constants $A$ and $B$ of \eqref{equ:SolCoreE} and \eqref{equ:SolCoreH}.
A convenient way to proceed is to calculate first the cylindrical-frame components of the fields:
%
\begin{align}
E^r:=&
   \frac{x E^x  +  y E^y} r = -\frac{i}{\psi^2k^2n^2-\beta^2}\left(\beta\partial_rE^z+\sqrt{\frac{\mu_0}{\epsilon_0}}\frac{\psi k}{r}\partial_{\theta}H^z\right)
    \,, \label{++equ:Er+}
\\
 E^{\theta}:=&
   \frac{x E^y  -{y} E^x} r =
 -
  \frac{i}{\psi^2k^2n^2-\beta^2}\left(\frac{\beta}{r}\partial_{\theta}E^z-\sqrt{\frac{\mu_0}{\epsilon_0}}\psi k \partial_rH^z\right)
  \,,
  \label{++equ:Etheta+}
\\
 H^r=&-\frac{i}{\psi^2k^2n^2-\beta^2}\left(\beta\partial_rH^z-\sqrt{\frac{\epsilon_0}{\mu_0}}\frac{\psi k n^2}{r}\partial_{\theta}E^z\right)
 \,,
   \label{++equ:Hr+}
\\
 H^{\theta}=&-\frac{i}{\psi^2k^2n^2-\beta^2}\left(\frac{\beta}{r}\partial_{\theta}H^z+\sqrt{\frac{\epsilon_0}{\mu_0}}\psi k n^2 \partial_rE^z\right)
  \,.
  \label{++equ:Htheta+}
\end{align}
The vanishing of surface currents imposes the requirement of continuity of  $E^{\theta}$ and $H^{\theta}$ at the core-cladding interface.%
\footnote{We check below (cf.\ \eqref{5III18.11}-\eqref{5III18.11asdf}) that the requirement of continuity of $B^r$, as needed for non-existence of magnetic charges, is equivalent to continuity of $E^\theta$. We also note that the non-occurrence of surface charges, which is equivalent to continuity of $D^r$, turns out to be equivalent to continuity of $H^\theta$.
}

We insert our solutions for $E^z$ and  $H^z$  into \eqref{++equ:Etheta+} and \eqref{++equ:Htheta+}, which results for $r=a$ in
%
\begin{align}
&
 \frac{1}{U^2}\left(im\beta A-\sqrt{\frac{\mu_0}{\epsilon_0}}\psi k B U \frac{J'_m(U)}{J_m(U)}\right)
 +
  \frac{1}{W^2}\left(im\beta A-\sqrt{\frac{\mu_0}{\epsilon_0}}\psi k B W \frac{K'_m( W )}{ K_m(W) }\right)=0 \label{equ:pPara}
\\
 &\frac{1}{U^2}\left(im\beta B+\sqrt{\frac{\epsilon_0}{\mu_0}}\psi k n_1^2 A U \frac{J'_m(U)}{J_m(U)}\right)
   +
  \frac{1}{W^2}\left(im\beta B
   +
   \sqrt{\frac{\epsilon_0}{\mu_0}}\psi k n_2^2 A W \frac{ K'_m(W) }{ K_m(W) }\right)=0
    \label{19VI18.1}
\end{align}
where the derivative is with respect to the argument, i.e. $J'_m(U)=\frac{dJ_m(U)}{dU}$.

We can also write \eqref{equ:pPara}-\eqref{19VI18.1} in matrix form: 
%
\begin{equation}
 \left(\begin{matrix}
im\beta \left(\frac{1}{U^2}+\frac{1}{W^2}\right) & -\sqrt{\frac{\mu_0}{\epsilon_0}}\psi k  \left(\frac{1}{U}\frac{J'_m(U)}{J_m(U)}+\frac{1}{W}\frac{ K'_m(W) }{ K_m(W) }\right) \\
\sqrt{\frac{\epsilon_0}{\mu_0}}\psi k \left(\frac{n_1^2}{U}\frac{J'_m(U)}{J_m(U)}+\frac{n_2^2}{W}\frac{ K'_m(W) }{ K_m(W) }\right) & im\beta \left(\frac{1}{U^2}+\frac{1}{W^2}\right)
\end{matrix}\right) \left(\begin{matrix}
A \\ B
\end{matrix}\right)=0
 \,.
  \label{1III18.1}
\end{equation}
%

For a solution other than $A=B=0$ we need a vanishing determinant, which results in
\begin{equation}
\frac{m^2\beta^2}{ \psi^2 k^2}\left(\frac{1}{U^2}+\frac{1}{W^2}\right)^2
 =
 \left(\frac{1}{U}\frac{J'_m(U)}{J_m(U)}
 +\frac{1}{W}\frac{ K'_m(W) }{ K_m(W) }\right)
  \left(\frac{n_1^2}{U}\frac{J'_m(U)}{J_m(U)}+\frac{n_2^2}{W}\frac{ K'_m(W) }{ K_m(W) }\right)
  \,.
\label{equ:ModeEqu}
\end{equation}
This is the key equation that determines $\beta$ as a function of $\phi$. While it cannot be solved in an explicit form using elementary functions, its solution gives the exact formula for $\beta$ in a step index fiber in terms of the remaining data.

In order to get insight into \eqref{equ:ModeEqu}, we use the definitions of $U$ and $W$, equations \eqref{equ:U} and \eqref{equ:W}, to get the relations
\begin{align}
 \frac{\beta^2}{k^2}= {\psi}^2n_1^2-\frac{U^2}{a^2k^2}
  \,,
\\
 \frac{\beta^2}{k^2}= {\psi}^2n_2^2+\frac{W^2}{a^2k^2}
 \,.
\end{align}
Dividing the first equation by $U^2$, the second by $W^2$, and adding the results  gives the relation
\begin{equation}
 \frac{\beta^2}{\psi^2k^2}\left(\frac{1}{U^2}+\frac{1}{W^2}\right)
 =\left(\frac{n_1^2}{U^2}+\frac{n_2^2}{W^2}\right)
 \,.
 \label{7III18.1abc}
\end{equation}

We note that the requirement of continuity of $B^r = \mu_0 H^r$ reads
\begin{equation}\label{5III18.11}
  i\beta\left( \frac 1 W \frac{K_m'(W)}{K_m(W)} + \frac 1 U  \frac{J_m'(U)}{K_m(U)}\right) B +
  \sqrt{\frac{\epsilon_0}{\mu_0}}\psi k m \left(\frac{n_1^2}{U^2}+\frac{n_2^2}{W^2}\right) A = 0
  \,,
\end{equation}
which by \eqref{7III18.1abc} is equivalent to
\begin{equation}\label{5III18.11asdf}
  i\sqrt{\frac{\mu_0}{\epsilon_0}}\left( \frac 1 W \frac{K_m'(W)}{K_m(W)} + \frac 1 U  \frac{J_m'(U)}{K_m(U)}\right) B
   +
 \frac{\beta m }{  \psi k} \left(\frac{1}{U^2}+\frac{1}{W^2}\right) A = 0
  \,.
\end{equation}
Comparing with the second line of~\eqref{equ:pPara}, we see that continuity of $E^\theta$ is equivalent to that of $B^r$. As already pointed out, one checks similarly that continuity of $H^\theta$ turns out to be  equivalent to continuity of $D^r$.

Inserting \eqref{7III18.1abc} into \eqref{equ:ModeEqu} and dividing both sides by $n_1^2$ gives
%
\begin{eqnarray}
\nonumber
 \lefteqn{
 m^2\left(\frac{1}{U^2}+\frac{1}{W^2}\right)\left(\frac{1}{U^2}
  +\frac{n_2^2}{n_1^2}\frac{1}{W^2}\right)
  =
  }
  &&
\\
   &&
    \left(\frac{1}{U}\frac{J'_m(U)}{J_m(U)}+\frac{1}{W}\frac{ K'_m(W) }{ K_m(W) }\right)\left(\frac{1}{U}\frac{J'_m(U)}{J_m(U)}+\frac{n_2^2}{n_1^2}\frac{1}{W}\frac{ K'_m(W) }{ K_m(W) }\right)
 \,,
\label{equ:ModeEquN}
\end{eqnarray}
independently of $\psi$
when $U$ and $W$ are viewed as independent variables. We write this equation in the symbolic form
\bel{13X17.1}
 F_m(U,W)=0
 \,,
\ee
and remark that $\beta$ enters this equation through $U$ and $W$ only.
Further, neglecting momentarily the dependence of the physical radius of the core upon $\phi$ and of $n$ upon the wavelength, the gravitational potential
$\phi$ enters \eqref{13X17.1} only through  the dependence of $U$ and $W$ upon
\begin{equation}\label{13X17.2}
  \mbox{$ \bar k:= \psi k$.
  }
\end{equation}

Now, we are interested in the change of $\beta$, say $\delta \beta$, after a change of height $\delta h$, for small $\delta h$. One can normalise all fields of interest so that $\phi=0$ at $h=0$, and approximate
\begin{equation}\label{27IX17.11}
  \beta\approx \beta|_{\phi =0 }
   +
    \left(\frac{\partial \beta}{\partial \phi} \frac{\partial \phi}{\partial h}\right)\Big |_{\phi =0 }
     \delta h
  =: \beta_0 + \beta'  g_{\earth}  \delta h =: \beta_0 + \delta \beta
  \,.
\end{equation}
(Recall that we use the symbol $ g_{\earth} $
to denote the gravitational acceleration at the surface of the earth.)
Differentiating \eqref{13X17.1}, we obtain
%
\begin{equation}\label{27IX17.13+}
 \Big( \partial_U  F_{m } \frac{\partial U}{\partial \beta}  +
  \partial_W  F_{m } \frac{\partial W}{\partial \beta}\Big) \frac{\partial \beta}{\partial \phi}
   +  \Big( \partial_U  F_{m } \frac{\partial U}{\partial \bar k}  +
  \partial_W  F_{m } \frac{\partial W}{\partial \bar k }\Big) \frac{\partial \bar k }{\partial \phi}
  = 0
  \,.
\end{equation}
Using ${\partial \bar k }/{\partial \phi}|_{\phi=0} = - 2 c^{-2} k   $, together with
\begin{equation}\label{13X17.4}
   \frac{\partial U}{\partial \beta} = - \frac{ a^2\beta} U
    \,,
    \quad
   \frac{\partial W}{\partial \beta} =    \frac{ a^2\beta}  W
    \,,
    \quad
   \frac{\partial U}{\partial \bar k } =  \frac { a^2n_1^2 \bar k} U
    \,,
    \quad
   \frac{\partial W}{\partial \bar k} =   - \frac { a^2n_2^2 \bar k}   W
    \,,
    \quad
\end{equation}
we can rewrite \eqref{27IX17.13+} as
\begin{equation}\label{27IX17.13+2}
 \Big( U^{-1}\partial_U  F_{m }   -
   W^{-1}\partial_W  F_{m }  \Big)\beta  \frac{\partial \beta}{\partial \phi}
   =
    -2 c^{-2} \Big(n_1^2 U^{-1} \partial_U  F_{m }-
 n_2^2 W^{-1} \partial_W  F_{m } \Big)   k ^2
  \,.
\end{equation}
Equivalently,
\begin{equation}\label{27IX17.14+}
  \fbox{$
   \displaystyle
   \delta \beta
    =
     - 2 \frac{
   n_1^2 W \partial_U  F_{m }-  n_2^2  U  \partial_W  F_{m } }
   { W\partial_U  F_{m }   -    U\partial_W  F_{m }  }
   \Big|_{\psi=1}
   \frac{  g_{\earth} k  ^2}{c^2 \beta} \delta h \,.$
  }
\end{equation}
Equation~\eqref{27IX17.14+} allows one to determine  $\delta \beta$ after solving numerically \eqref{equ:ModeEqu}.

As an example, consider a  waveguide with

\bel{1XI17.11}
 n_1 = 1.4712
 \,,
 \
 n_2 =  1.4659
 \,,\
 a =  4.1 \times {10^{-6} \mathrm{m}}
 \,,\
 k 
  \approx 4.05367 \times 10^6 {\mathrm{m}^{-1}}
 \,,
\ee
where $\mathrm{m}$ stands for meters, not to be confused with the parameter $m $ in \eqref{equ:SolCoreE}, \eqref{equ:SolCoreH} and elsewhere in the paper.
The numerical solution of \eqref{equ:ModeEqu} with $\delta h $ equal to one meter and with the radial order
parameter $m $  equal to one%
\footnote{We are very grateful to Maciej Maliborski for carrying   out the  {\sc Mathematica} calculation to the accuracy needed, and for generating Figure~\ref{F2III18.1}. The calculation turns out to be unreliable unless performed at very high precision.
\label{foot2III18.1}}
reads
\begin{equation}\label{31I18.1}
  \beta \approx 5.951705634994889611
  \times 10^6 \,\meter^{-1}
   \,.
\end{equation}

Comparing with the value at $\delta h=0$, the change of $\beta$ which results from elevating the waveguide by one meter leads to   a change of $\beta$ equal to
\bel{2XI17.12asd}
 \delta \beta \approx  - 1.301 \times 10^{-9} \meter^{-1}
 \,,
\ee
calculated either from \eqref{27IX17.14+}, or directly from \eqref{equ:ModeEqu}, with the difference between the values of $\delta \beta$ occurring only at the fourteenth significant digit.

\subsection{Variations of the diameter of the core}
 \label{24VI18}
We continue to ignore the fact that $n_1$ and $n_2$ depend upon $\phi$, and repeat the calculations above by taking into account that $a$ does. Differentiating again \eqref{13X17.1}, instead of \eqref{27IX17.13+} we obtain now
%
\begin{eqnarray}
\lefteqn{
 \Big( \partial_U  F_{m } \frac{\partial U}{\partial \beta}  +
  \partial_W  F_{m } \frac{\partial W}{\partial \beta}\Big) \frac{\partial \beta}{\partial \phi}
  }
  &&
   \nonumber
\\
 &&
   +  \Big( \partial_U  F_{m } \frac{\partial U}{\partial \bar k}  +
  \partial_W  F_{m } \frac{\partial W}{\partial \bar k }\Big) \frac{\partial \bar k }{\partial \phi}
   +  \Big( \partial_U  F_{m } \frac{\partial U}{\partial a}  +
  \partial_W  F_{m } \frac{\partial W}{\partial a}\Big) \frac{\partial a}{\partial \phi}
  = 0
  \,.
  \label{27IX17.13+n}
\end{eqnarray}
Using  ${\partial a }/{\partial \phi}|_{{\psi=1}} = c^{-2}  a_p|_{{\psi=1}}  = c^{-2} a $, together with
\begin{eqnarray}\label{13X17.4n}
&
 \displaystyle
   \frac{\partial U}{\partial a} = \frac U a
    \,,
    \quad
   \frac{\partial W}{\partial a} =    \frac   W a
    \,,
    &
\end{eqnarray}\label{13X17.4Add}
we  rewrite \eqref{27IX17.13+n} as
\begin{eqnarray}
\lefteqn{
 \displaystyle
 c^2 \Big( U^{-1}\partial_U  F_{m }   -
   W^{-1}\partial_W  F_{m }  \Big)\beta  \frac{\partial \beta}{\partial \phi}
   }
   &&
    \nonumber
\\
 &&
 \displaystyle
   =
    -2  \Big(n_1^2 U^{-1} \partial_U  F_{m }-
 n_2^2 W^{-1} \partial_W  F_{m } \Big)   k ^2
  +
      \Big( U  \partial_U  F_{m }
      +
 W  \partial_W  F_{m } \Big)   a ^{-2}
  \,.
   \phantom{xxxx}
\end{eqnarray}\label{27IX17.13+2n}
Instead of \eqref{27IX17.14+} we thus have 
\begin{eqnarray}
   \displaystyle
   \delta \beta
   & \equiv
    &
    (\delta \beta)_k + (\delta \beta)_a
\nn
\\
 &
    = &
     \frac{- 2 \Big(
   n_1^2 W \partial_U  F_{m }-  n_2^2  U  \partial_W  F_{m }
           \Big)
            k  ^2
    +
      UW \Big( U  \partial_U  F_{m }
      +
 W  \partial_W  F_{m } \Big)   a ^{-2}}
   { W\partial_U  F_{m }   -    U\partial_W  F_{m }  }
   \Big|_{\psi=1}
   \frac{   g_{\earth} }{c^2 \beta} \delta h \,.
\nn
\\
 &&
    \label{27IX17.14+n}
\end{eqnarray}
%
where $(\delta \beta)_k$ is the change of $\beta$
arising from the $k^2$-term in the second line of~\eqref{27IX17.14+n}, already seen in \eqref{27IX17.14+}.

The contribution
from the $a^{-2}$-term in ~\eqref{27IX17.14+n},  denoted by $(\delta \beta)_a$, can be calculated to be
\bel{2XI17.12asd2as}
 (\delta \beta)_a \approx 1.507 \times 10^{-12} \meter^{-1}
 \,,
\ee
thus by about three orders of magnitude smaller than \eqref{2XI17.12asd}.

\subsection{$\lambda$-dependent index of refraction}
 \label{s24VI18.11}

We consider finally the effect arising from the dependence of the indices of refraction $n_1$ and $n_2$ upon the \emph{vacuum} wavelength $\lambda_p$ of light,
$$
 \lambda_p:=  c T_p \equiv \frac{2 \pi c}{\omega_p}
 \,,
$$
where $T_p$ is the proper-time period of the wave and $\omega_p$ its proper-time frequency. The latter is related to the coordinate frequency $\omega$ as
$$
 \omega_p= \frac{\omega}{\sqrt{ 1 + 2 c^{-2} \phi}}
   \approx \omega \left(1 - \frac {\phi}{c^2}\right)
 \,.
$$
Thus
\begin{equation}\label{24VI18.1}
  \lambda_p \approx \frac {2 \pi }{k} \left(1 + \frac {\phi}{c^2}\right)
   \,.
\end{equation}

We will use Sellmeier's formula~\cite{Sellmeier}

\begin{equation}\label{24VI18.2}
  n (\lambda_p) =
   \sqrt{C_0 + \sum_{i=1}^{N} \frac{c_i \lambda_p^2}{
    {\lambda_p^2 - k_i}
    }
    }
    \,,
\end{equation}
with $N=3$  where, for pure silica glass,
the phenomenological constants are~\cite[p.~92]{Okamoto}  $C_0=1 $ and
%
\begin{eqnarray}
 \nonumber 
    &
 c_1= 0.6965325
 \,,
 \quad
 k_1=(6.60932   \times  10^{-8}\mathrm{m})^{2}
 \,,
 &
   \\
    &
    \nonumber
 c_2=0.4083099
 \,,
 \quad
  k_2=(1.1811 \times  10^{-7}\mathrm{m}^{ })^2
 \,,
 &
   \\
    &
  c_3= 0.8968766
 \,,
 \quad
   k_3=
    (9.89616  \times  10^{-6} \mathrm{m}^{})^2
 \,.
 &
\end{eqnarray}
Hence
\begin{equation}\label{24VI18.3}
  \frac{\partial n}{\partial \phi}\Big|_{\psi=1}
  = \frac{2 \pi}{ k c^2}
  \frac{\partial n}{\partial \lambda_p}\Big|_{\psi=1}
   \,,
\end{equation}
with
\begin{equation}\label{25VI18.1}
  \frac{\partial n}{\partial \lambda_p}\Big|_{\psi=1}
   = - \frac{  \lambda_p }{ n  }\sum_{i=1}^N  \frac{c_ik_i}{(\lambda^2_p-k_i)^2}
   \approx
  -\frac{1.7285 \times 10^4}{n}
  \times \meter^{-1}
   \,.
\end{equation}

Adapting the calculations so far to take into account the dependence of $n$ upon $\phi$, and setting $n=(n_1+n_2)/2$, one finds in the right-hand side of \eqref{27IX17.14+n} a supplementary term
%
%
\begin{eqnarray}
   (\delta \beta)_{\lambda_p}
    &:= &
     \frac{ 
   n_1 W \partial_U  F_{m }\frac{\partial n_1}{\partial \lambda_p} -  n_2  U  \partial_W  F_{m }\frac{\partial n_2}{\partial \lambda_p}
            }
   { W\partial_U  F_{m }   -    U\partial_W  F_{m }  }
   \Big|_{\psi=1}
   \frac{ 2 \pi  k     g_{\earth} }{c^2 \beta}
   \delta h
   \nonumber
\\
    &
    \approx
     &
     \underbrace{
     \frac{ 
   n_1 W \partial_U  F_{m }-  n_2  U  \partial_W  F_{m }
            }
   { W\partial_U  F_{m }   -    U\partial_W  F_{m }  }
   \Big|_{\psi=1}
   }_{\approx 1.46992}
    \frac{\partial n}{\partial \lambda_p} \Big|_{\psi=1}
   \underbrace{
   \frac{ 2 \pi  k     g_{\earth} }{c^2 \beta}
   }_{\approx 4.65977 \times 10^{-16}\mathrm{m}^{-1}} \delta h
   \nonumber
\\
  &
   \sim
   &10^{-11} \mathrm{m}^{-2} \delta h
   \,,\label{27IX17.14+lp}
\end{eqnarray}
which is two orders of magnitude smaller than the dominant effect $(\delta \beta)_k$.

\subsection{Change of phase}
 \label{s24VI18.1}

Summarising, we have shown that the overall effect is, for small $\delta h$,
\bel{2XI17.12asd2as2}
(\delta \beta)_k +  (\delta \beta)_a
  + (\delta \beta)_{\lambda_p}
\approx (\delta \beta)_k
  \approx  - 1.30 \times 10^{-9} \meter^{-1}\times  \delta h
 \,.
\ee

Consider two identical waveguides extended horizontally, of physical length $L_p$ and therefore coordinate length
$$
 L=\psi^{-1/2} L_p
  \,,
$$
located at heights differing by $\delta h$, with coordinates normalised so that $\psi=1$ at the location of the lower waveguide. We assume that the coordinate frequency $\omega$ is the same for both waveguides, which will be the case if the incoming light comes from the same laser. The phase difference between the light arriving at the far end of the waveguides will be, in the linear approximation, taking $L_p$ to be $10^5 \mathrm{m}$,
\bean
 \delta (\beta  \times \psi^{-1/2}  L_p )|_{{\psi=1}}
  & =  &
  \big(
   \delta  \beta|_{{\psi=1}}   +
    \underbrace{\frac { g_{\earth}}{c^2} \beta }_{\sim 6.487 \times 10^{-10} \mathrm{m}^{-2}}
     \delta h \big)
 \times   L \approx
  \delta  \beta|_{{\psi=1}}  \times   L
\\
 &
    \approx
 &
    -6.52 \times 10^{-5}  \meter^{-1} \times \delta h
 \,.
\eeal{2XI17.12ase}
An experiment, with one arm of the interferometer  in the basement of the Bremen ZARM tower and the other at the top, would thus observe a phase shift of the order of $10^{-3}$ radians.

\subsection{Weakly guiding approximation}
 \label{ss30I18.1}

In most practical fibers used, the difference of the refractive indices $n_1$ and $n_2$ is small, so we can assume $n_1 \simeq n_2=:n$ in some relations. This is usually referred to as the \emph{weakly guiding approximation}, due to the fact that the waveguide is almost homogeneous when $n_1$ almost equals $n_2$.
Taking the limit $n_1\to n_2 \approx n $ in
\eqref{27IX17.14+} one obtains
\begin{equation}\label{28IX17.1}
  \fbox{$
   \displaystyle
    \delta \beta  \approx (\delta \beta)_k =   - \frac{2  k^2 n^2   }{ c^2 \beta}  \delta \phi
   = -  \frac{2 \omega^2 n^2 g }{ c^{ 4} \beta}  \delta h \,.$
  }
\end{equation}
If we use the parameters
\eqref{1XI17.11}, and set again $n= (n_1+n_2)/2$,
\eqref{28IX17.1} becomes
\begin{equation}\label{11XI17.13}
  \fbox{$
   \displaystyle
   (\delta \beta)_k =  -1.297   \times 10^{-9} \meter^{-2}\times \delta h \,,$
  }
\end{equation}
which agrees reasonably well with the linear approximation to the exact result \eqref{2XI17.12asd}.

The simplicity of \eqref{28IX17.1} might appear surprising at first sight, but its derivation is actually much more straightforward than the calculations so far. Indeed, since
$$
 \psi n_2 k <  \beta < \psi n_1 k
 \,,
$$
the weakly guiding limit $n_1, n_2 \to n$ gives
\begin{equation}\label{16X17.1}
 \beta = \psi n  k
 \,,
\end{equation}
and \eqref{28IX17.1} immediately follows.


\section{Gravitational potential varying along the waveguide}
 \label{s10II18.1} 

To account for a non-constant gravitational potential we start with the full form \eqref{27VII17.3}-\eqref{27VII17.4} of the  post-Newtonian Maxwell equations.
Setting
\begin{equation}\label{1II18.3}
  H^i = \mu_0^{-1} \blueB^i
  \,
\end{equation}
we have to analyze the following system of equations:
\begin{align}
\psi \mu_0\partial_t\vec{H}(\vec{r},t)&=-\nabla \times \vec{E}(\vec{r},t) -\frac{2}{c^2}\left(\nabla\phi \times \vec{E}(\vec{r},t)\right) \label{pot:Fara}
 \,,
\\
\psi \epsilon_0n^2(r)\partial_t\vec{E}(\vec{r},t)&=\nabla \times \vec{H}(\vec{r},t) +\frac{2}{c^2}\left(\nabla\phi \times \vec{H}(\vec{r},t)\right) \label{pot:Max}
 \,,
\\
\nabla \cdot \vec{E}(\vec{r},t)&=0 \label{pot:divE}
 \,,
\\
\nabla \cdot \vec{H}(\vec{r},t)&=0 \label{pot:divH}
 \,.
\end{align}

Consider the second-order wave equation \eqref{27VII17.3asdf}.
Estimating $ |\mathring \Delta \vec B|$ by $ | \vec B|\lambda^{-2}$, $|\nabla \phi|$ by $g_{\earth}$, $|\nabla \nabla \phi|$ by ${g_{\earth}/R_{\earth} \mdanger}$, where $\lambda$ is the wavelength of the photon in the waveguide,
we find
that 
\begin{equation}\label{16VIII17.2}
  |\nabla \phi| c^{-2} |\nabla \vec B |\approx \fbox{$10^{-22}$} |\mathring \Delta \vec B|
  \,,
  \quad
  |\nabla \nabla \phi| c^{-2} | \vec B |\approx \frac{\lambda }{ R_{\earth}  } |\nabla \phi| c^{-2} |\nabla \vec B |
  \approx \fbox{$ 10^{-35}$} | \mathring \Delta \vec B|
   \,.
\end{equation}
Thus the corrections arising from second derivatives of $\phi$ are much smaller than those coming from the gradient-$\phi$ terms.
We therefore neglect the second derivatives of $\phi$ in \eqref{27VII17.3asdf}, obtaining the following wave equations for the Cartesian components  $H^i$ of  $\vec H$:
\begin{eqnarray}
 \label{1II18.4}
 &
 \left(
  1- \frac{2 \phi}{  c^2}
  \right)^2
   \epsilon \mu
   \partial_{t}^2 H^k =   \mathring \Delta H^k
   + \frac 1 {c^2} \partial^i \phi(
   6\partial_i  H^k -
   4 \partial_k H^i)
 \,,
 &
\end{eqnarray}
where $\mathring \Delta$ is the flat-space Laplace operator $\partial_x^2 + \partial_y^2 + \partial_z^2$. The Cartesian components of $\vec E$ satisfy an identical equation.
Here $\phi$ is the gravitational potential normalised to zero at some chosen reference point in the lab, compare~\eqref{19X17.1+}.

We want to determine the change of $\beta$ that arises in a coil of  {radius} $d_s\approx 10^{-1}m$  with axis of symmetry aligned with the horizontal direction in the lab.  We consider the most common geometrical arrangement for optical fibers, which is that of a cylindrical fiber spool of length $\ell$. As a first approximation we consider a rigid spool in Euclidean space.
The total  Euclidean  length of the fiber is denoted by $L= 2\pi N_s d_s$.

It is convenient to distinguish between the gravitational potential at a reference point in the lab, which we denote by $\mathring \phi$, and the gravitational potential on the axis of the spool, which we will denote by $\phi_0$.

When the difference between points on the waveguide and the reference point in the lab is at most of  the order of meters we have, by \eqref{19X17.1+},
\begin{equation}\label{6III18.1}
  \psi = 1 - \underbrace{\frac{2(\phi-\mathring \phi )}{c^2}}_{O(10^{-16})} + O (10^{-25})
  \,.
\end{equation}

A comment about the validity of the first post-Newtonian approximation is in order. Since $ {2 \phi}  c^{-2} \approx  - 1.39 \times 10^{-9}$ at the surface of the earth, the second-order Newtonian approximation introduces further terms in the Maxwell equations   which are expected to be of relative size $ ({2 \phi}  c^{-2})^2 \approx     10^{-18}$. This seems to be much bigger than the first-derivative correction as estimated in \eqref{16VIII17.2}, and the effect of which we are about to determine. However, when comparing the results of the calculations at different heights $h$, the relevant effect on the equations is not that arising from the actual values of the metric components at the surface of the earth, but from their gradients, which will be smaller by a factor $1/R_{\earth}$, hence a contribution $O(10^{-24})$ from the second-order post-Newtonian corrections, and thus smaller than \eqref{16VIII17.2}.

It appears rather daunting to try to find an explicit solution of the Maxwell equations describing the above geometry. A simplified model is therefore needed.
To proceed, let $s$ be a Euclidean-length parameter along a waveguide.  We choose the $z$-axis to be aligned with the axis of the spool,
the waveguide is then described as the following curve in $\mathbbm{R}^3$:
\begin{equation}\label{8X17.1}
  [0,L]\ni s \mapsto
  \Big( d_s \cos \big(\frac{s-s_0}{ d_s } \big), d_s\sin \big(\frac{s-s_0}{ d_s } \big),\frac{\ell}{L}s\Big)
   \,,
\end{equation}
where $s_0$ describes the entry-point of the waveguide.
If we take the $x$-axis to be the vertical, the  (suitably normalised)  gravitational potential at a point of the waveguide parameterised by $s$ is equal to
\begin{equation}\label{8X17.3}
  \phi_0  {-\mathring \phi}+ g_{\earth}   d_s \cos \big(\frac{s-s_0}{ d_s }\big)
   \,.
\end{equation}

We now ``unwind'' the waveguide to a straight-one  lying along the $z$-axis, replacing the parameter $s$ in \eqref{8X17.3} by $z$, and think of the potential \eqref{8X17.3} as a function along the waveguide:
\begin{equation}\label{16VIII17.1}
  \phi(z)= \phi_0  {-\mathring \phi} + {g_{\earth} d_s \mdanger}   \cos \big(\frac{z-z_s}{ d_s } \big)
  \,.
\end{equation}

The question then arises, how to handle the gradient-$\phi$ terms in \eqref{1II18.4} after this ``unwinding''. By inspection of orders of magnitude, as just discussed above as well as in Section~\ref{ss3VIII17.1}, or \emph{a posteriori} by inspection of the final calculations, one finds that the dominant effect of a non-constant $\phi$ on the solution comes from the undifferentiated term in \eqref{1II18.4}.

In any case,  the components of the gradient terms in the equations in directions transversal to the fibre mix the tangential and the transverse components of the Maxwell fields, which complicates  considerably the analysis without affecting the dominant, already small, effect. On the other hand, tangential gradients do not introduce any supplementary complications. Thus, for the sake of a future more detailed analysis, and as a tool for cross-checking the subdominant effect of the gradient terms, we will keep the terms arising from the $z$-gradient of $\phi$ in our calculations,  ignoring the transverse ones.

In view of the above, 
using the time-periodic ansatz \eqref{29I18.2}-\eqref{1II18.7+} in \eqref{1II18.4} leads to
%
\begin{equation}
-\psi^2n^2k^2H^z= \mathring \Delta H^z+\frac{2}{c^2} \partial_z\phi\partial_zH^z
 \,.
  \label{1II18.7}
\end{equation}

The equation for the $z$-component of the electric field is identical.

Let us introduce a small dimensionless constant $ \varepsilon $ by setting
\begin{equation}\label{19I18.3}
   \varepsilon =\frac {g_{\earth }d_s} {c^2} \sim 10^{-17} \ll 1
  \,,
\end{equation}
and let
\begin{equation}\label{19I18.3+}
  \beta_1 := d_s^{-1} \sim 10 \, {\mathrm{m}^{-1}}
  \ll \beta \sim 10^6 {\mathrm{m}^{-1}}
  \,,
  \quad
 {\aone}:=  1-\frac{ {2(\phi_0-\mathring \phi)}}{c^2}
   \approx 1
 \,.
\end{equation}
These values will be assumed below whenever numerical estimates are used.

Shifting $z$ by $z_s$, and neglecting terms which are higher-order in $ \varepsilon $, \eqref{1II18.7} becomes
%
\begin{equation} \label{19I18.1}
 \partial^2_rH^z+\frac{1}{r}\partial_rH^z
  +\frac{1}{r^2}\partial^2_{\theta}H^z+\partial^2_zH^z
  +\big({\aonesquared }\plustwoatwo \varepsilon  \cos(\beta_1 z )\big) {k^2n^2} H^z
   {-}
2 \varepsilon
   {\noatwo} \sin(\beta_1 z )
   \partial_zH^z
  =0
   \,.
\end{equation}
We seek a solution of the form
\begin{equation}\label{19I18.2}
  H^z(r,\theta,z) = e^{i m \theta} \sum_{\ell \in \Z}
    \smallh^z_\ell (r) e^{i( \ell  \beta_1 {-\beta} )z }
  \,.
\end{equation}
Inserting in \eqref{19I18.1}, for each $\ell$ we obtain
%
\begin{eqnarray}
 \nonumber
 \lefteqn{
   \partial^2_r \smallh^z_\ell  +\frac{1}{r}\partial_r \smallh^z_\ell
  -\frac{m^2}{r^2}\smallh^z_\ell
  -  {(\beta-\ell \beta_1)^2} \smallh^z_\ell
  }
  &&
\\
 &=&
   (\beta-\ell \beta_1) \varepsilon
   {\noatwo} (\smallh^z_{\ell +1}- \smallh^z_{\ell -1})
  - {k^2n^2}
    \big({\aonesquared } \smallh^z_\ell  \plusatwo \varepsilon (\smallh^z_{\ell +1}+ \smallh^z_{\ell -1})\big)
 \nn
\\
 &&
  {-
 \varepsilon\beta_1^2\big(\smallh^z_{\ell +1}+ \smallh^z_{\ell -1}\big)}
   \,.
    \phantom{xx}
    \label{19I18.4b}
\end{eqnarray}

Some comments about the structure of the equations, and of the error terms, are in order.

First, since \eqref{1II18.7} is linear in $H^z$, we have the freedom to rescale $H^z$, and the influence of  all error terms on the solution scales accordingly.

Next, when passing from \eqref{1II18.7}
to \eqref{19I18.1} we have ignored terms of the order of $\varepsilon^2 |H^z|$, and thus all the equations that follow will not be more accurate than this. In other words, the right-hand side of \eqref{19I18.1} is not really zero but $O(\epsilon^2 |H^z|)$. Here, and in what follows, $f=O(h)$ means that there is a constant $C$  such that
$$
 |f| \le C |h|
\,.
$$
 \redad{18VI18}{
Care must be taken because the constant $C$ could depend  upon $\ell$ in the calculations that follow, and could grow without bound as $\ell$ increases. This issue needs to, and will be, addressed in our analysis below.
}

Similarly the right-hand side of \eqref{19I18.4b} should contain a supplementary term $O(\epsilon^2 |H^z|)$. Let us momentarily ignore this and take this equation at face value. Then \eqref{19I18.4b} forms an infinite system of coupled equations for the ``$\smallh_\ell^z$-modes'', where the mixing between different $\ell$'s involves a small constant $\varepsilon$. Anticipating a more detailed analysis that will be presented shortly,  a coefficient $\mathringB_0$, which will determine the size of $|H^z|$, can be chosen arbitrarily after a suitable choice of $\beta$. Next, each of the equations
\eqref{19I18.4b} is a PDE for $\smallh_\ell^z$ when the ``neighbouring'' fields $\smallh_{\ell\pm 1}^z$ are known. This implies that, after solving,
$\mathringB_0$ propagates to $\smallh_{\pm 1}^z$ as $\varepsilon $ times a factor, say $ {\mdanger \zeta} \ge 1$, which needs to be determined and which turns out to be large. Then
$\mathringB_0$  propagates to $\smallh_{\pm2}^z$ as $\varepsilon^2$ times another numerical factor, and so on. Assuming that the associated numerical factors are of the same order at each step, and denoting by
$$
 \varepsilon_1=  {\mdanger \zeta} \varepsilon
$$
the product of $\varepsilon$ with a typical factor arising in this rough argument,
and further assuming that $\varepsilon_1$ remains much smaller than one, one is led to expect that $\mathringB_0$ will give a contribution to $\smallh^z_\ell$ which is of order
$O(\epsilon_1^{|\ell|} |\mathringB_0|)$.

Now, viewing again the right-hand side of \eqref{19I18.4b} as known, there is a freedom of adding to $\smallh_\ell^z$ a solution of the homogeneous PDE. We will exploit this freedom to ensure continuity of the remaining fields of interest, with the precise value of the associated free coefficients determined at the end of the calculation. Hoping that the resulting contribution to $\smallh_\ell^z$ remains
$O(\varepsilon^{|\ell|}_1 |\mathringB_0|)$, we should end up with a solution such that
\begin{equation}\label{24III18.1}
  \smallh_\ell^z =
 O(\varepsilon^{|\ell|}_1  |\mathringB_0|)
 \,.
\end{equation}

We expect that the existence of solutions of \eqref{19I18.4b} satisfying \eqref{24III18.1} can be established rigorously, but we have not checked all details of the proof.

It turns out to be convenient to write $\smallh^z_\ell $ as
\begin{equation}\label{19I18.5}
  \smallh^z_\ell (r) = \mathring{\smallh}{}^z_\ell (r) +  \varepsilon  \delta \smallh^z_\ell (r)
  \,,
\end{equation}
where $\mathring{\smallh}{}^z_\ell $ satisfies
%
%
\begin{eqnarray}
   \big(
     \partial^2_r +\frac{1}{r}\partial_r -\frac{m^2}{r^2}
      \big)
      \mathring{\smallh}{}^z_\ell
 = \big(
    {(\beta-\ell \beta_1)^2}
   - {\aonesquared }  {k^2n^2}
  \big)
   \mathring{\smallh}{}^z_\ell
   \,.
    \label{19I18.8}
\end{eqnarray}

Then, ignoring again terms involving $ {\varepsilon ^2}$,
 the functions $\delta \smallh^z_
\ell$ are solutions of the equations
\begin{eqnarray}
\nonumber
 \lefteqn{
    \partial^2_r \delta \smallh^z_\ell  +\frac{1}{r}\partial_r \delta \smallh^z_\ell
  +\big( {k^2n^2}  {\aonesquared }
  -  {(\beta-\ell \beta_1)^2}
  -\frac{m^2}{r^2}
  \big) \delta \smallh^z_\ell
  }
  &&
\\
 &&
 =
     (\beta-\ell \beta_1)
    {\noatwo} ( \mathring{\smallh}{}^z_{\ell +1}-  \mathring{\smallh}{}^z_{\ell -1})
  +(2 {k^2n^2}   {- \beta_1^2 }
 )( \mathring{\smallh}{}^z_{\ell +1}+  \mathring{\smallh}{}^z_{\ell -1})
   \,.
    \phantom{xx}
    \label{19I18.4}
\end{eqnarray}

A detailed discussion of the terms neglected is again in order. Keeping in mind the term $O(\epsilon^2 |\mathring H^z|)=O(\epsilon^2 |\mathringB_0|) $ omitted in \eqref{19I18.4b},
 \redad{18VI18.1}{
as well as our expectation
\begin{equation}\label{21VI18.1}
 \delta \smallh_{\ell } ^z = O(\varepsilon^{-1}\varepsilon^{|\ell |}_1  |\mathringB_0|)
 \,,
 \quad |\ell |>0
 \,,
 \quad
 \delta \smallh_{0 } ^z = O(\varepsilon^{-1}\varepsilon^{2}_1  |\mathringB_0|)
\end{equation}
(compare \eqref{24III18.1}) and  the relation $\epsilon_1= {\mdanger \zeta}\varepsilon$, the terms ignored when passing from \eqref{19I18.4b} to the right-hand side of \eqref{19I18.4} are
%
\bel{24III18.11}
 \left\{
   \begin{array}{ll}
O(\varepsilon |\mathringB_0|) + O(k^2  \varepsilon_1 |\mathringB_0|), & \hbox{$\ell=0$;} \\
O(\varepsilon |\mathringB_0|) + O(k^2  \varepsilon_1^2 |\mathringB_0|), & \hbox{$\ell=\pm1$;} \\
O(\varepsilon |\mathringB_0|) + O(k^2\varepsilon^{|\ell|-1}_1 |\mathringB_0|), & \hbox{otherwise.}
   \end{array}
 \right.
\ee
For $|\ell|\ge 2$, arguing as before, we expect this to lead to
 an error in $\delta\smallh_\ell$  of order
$$
  {\mdanger \zeta} \big( O(\varepsilon |\mathringB_0|) + O(\varepsilon^{|\ell|-1}_1 |\mathringB_0|)\big)
 =
  O(\varepsilon_1 |\mathringB_0|) + O( {\mdanger \zeta}\varepsilon^{|\ell|-1}_1 |\mathringB_0|)
 \,,
$$
which is consistent with \eqref{24III18.1}. Here $ {\mdanger \zeta}$ needs to be chosen large enough to absorb  the contribution from $k^2$ in the error terms.

Similarly the error arising in $\smallh_0^z$ is
$$
  \varepsilon\big(O(\varepsilon_1 |\mathringB_0|) + O( {\mdanger \zeta} \varepsilon_1  |\mathringB_0|)\big)
=
  O(\varepsilon\varepsilon_1  |\mathringB_0|) + O(\varepsilon_1^2 |\mathringB_0|)\big)
 \,.
$$
A similar calculation applies for $\ell=\pm1$. This confirms consistency of the scheme, and also implies that no accuracy will be gained in going beyond $\ell=\pm 1$ in further calculations.
}

Equation~\eqref{19I18.8} can be solved as in \eqref{equ:SolCoreH}:
\begin{equation}
 \mathring{\smallh}{}^z_\ell(r)=  {\nomathringB}_\ell \mathring g_\ell(r)
\,,
\
\mbox{with}
 \
 \mathring g_\ell(r)=
\begin{cases}
  \displaystyle  \frac{J_m(U_\ell r/a)}{ J_m(U_\ell)  } \,,          & \text{core,}\\
  \displaystyle   \frac{K_m(W_\ell r/a)}{ K_m(W_\ell)  }  \,,              & \text{cladding,}
\end{cases}
 \label{19I18.9}
\end{equation}
where $ {\nomathringB}_\ell$ is a constant and
\begin{equation}
U^2_\ell:=a^2
 \big(
   {\aonesquared }  {k^2n^2_1}
  -  {(\beta-\ell \beta_1)^2}
   \big)
   \label{19I18.10a}
 \,,
\end{equation}
\begin{equation}
W^2_\ell:=a^2
    \big(
     {(\beta-\ell \beta_1)^2} -  {\aonesquared }  {k^2n^2_2}
     \big)
\,,
 \label{19I18.10}
\end{equation}
provided that the right-hand sides of \eqref{19I18.10a}-\eqref{19I18.10} are positive, and that the denominators in \eqref{19I18.9} do not vanish. 
(Thus, the constants $A_0$ and $B_0$  here correspond to the constants $A$ and $B$ of  Section~\ref{ss3VII18.1}.) 
Using \eqref{19I18.8}, a particular solution of  \eqref{19I18.4} is obtained by setting
\begin{equation}\label{20I19.1}
  \delta \smallh^z_\ell =  {\mdanger \zeta}_{\ell,-} \mathring{\smallh}{}^z_{\ell-1}
  +
   {\mdanger \zeta}_{\ell,+} \mathring{\smallh}{}^z_{\ell+1}
   \,,
\end{equation}
with
%
\begin{equation}\label{20I19.2}
    {\mdanger \zeta}_{\ell,-}=
\frac{\big( {-\beta}+(\ell-1) \beta_1\big) {\noatwo}  {+} {2}  {k^2n^2}}
   {\beta_1(  {2\beta}  {-}  (2\ell -1) \beta_1)}
   \,,
   \quad
   {\mdanger \zeta}_{\ell,+}
  =  \frac{
 \big( {-\beta}+(\ell+1) \beta_1\big)\beta_1
        {-}  {2k^2n^2} }
   {\beta_1(  { 2\beta}- (2\ell +1) \beta_1)}
   \,,
\end{equation}
assuming again that the denumerators do not vanish.
Using the values of the parameters \eqref{1XI17.11}-\eqref{31I18.1} and \eqref{19I18.3+} we find for small $\ell$
\begin{equation}\label{20I19.2+}
    {\mdanger \zeta}_{\ell,-}\approx
  - {\mdanger \zeta}_{\ell,+}\approx
\frac{  {k^2n^2} }
   {\beta_1   { \beta}}\sim 2.761 \times 10^5 n^2
\
 \Rightarrow
 \
 \varepsilon    {\mdanger \zeta}_{\ell,\pm}  \approx  \mp 6.514 \times 10^{-11} d_s \left(\frac{n}{n_1}\right)^2
   \,.
\end{equation}

In view of \eqref{20I19.2+}, we can choose
\begin{equation}\label{24III13.31}
  \varepsilon_1\sim 10^{-10}
 \,.
\end{equation}
Returning to our equations, ignoring terms involving $\varepsilon_1^2 $ we find
\begin{equation}\label{20I18.3}
  \smallh^z_\ell(r) =
  \left\{
    \begin{array}{ll}
        {\mathringB}_0 \mathring g_0(r), & \hbox{$\ell=0$;} \\
       \varepsilon
       \big(
         {\mathringB}_{1 } \mathring g_{ 1}(r) +  {\mdanger \zeta}_{1,-}  {\mathringB}_0 \mathring g_0(r)
         \big)
          , & \hbox{$\ell=1$;} \\
       \varepsilon
      \big(  {\mathringB}_{-1}
     \mathring g_{-1}(r) +  {\mdanger \zeta}_{-1,+} {\mathringB}_0 \mathring g_{0}(r)
      \big)
      , & \hbox{$\ell=-1$,}
    \end{array}
  \right.
\end{equation}
(so that $\nomathringB_{0} =   \mathringB_{0}$,  $\nomathringB_{\pm1} = \varepsilon \mathringB_{\pm 1}$)
with
\begin{equation}\label{9III18.11}
  | {\mathringB}_{ \pm 1}|   \varepsilon
\le  {\varepsilon_1} | {\mathringB}_{0}|
\,,\quad
 |   {\mdanger \zeta}_{\pm 1,\mp}| 
  \varepsilon
 \le  {\varepsilon_1}
 \,.
\end{equation}
We have also assumed that
\begin{equation}\label{20I18.1}
\forall \   \ell \in \{0,\pm 1\} \qquad
J_m(U_\ell)\neq 0
 \,,
 \
K_m(W_\ell)\ne 0
  \,,
 \
  { 2\beta} +  \ell \beta_1 \ne 0
  \,,
 \
   {\aonesquared }  {k^2n^2_2}  < ( \beta +\ell \beta_1)^2 <  {\aonesquared }  {k^2n^2_1}
  \,.
\end{equation}
Equation~\eqref{20I18.3} can be streamlined by setting
\begin{equation}\label{7III18.2}
   {\mdanger \mathring{\zeta}}_\ell =
 \left\{
  \begin{array}{ll}
    0, & \hbox{$\ell=0$;} \\
     {\mdanger \zeta}_{1,-}, & \hbox{$\ell=1$;} \\
     {\mdanger \zeta}_{-1,+}, & \hbox{$\ell=-1$,}
  \end{array}
 \right.
\end{equation}
whence,
\begin{equation}\label{20I18.3+}
 \mbox{for $\ell\in \{0,\pm 1\}$, }
 \
 \smallh^z_\ell(r) =
 \varepsilon^{|\ell|}
 \left(
         {\mathringB}_{\ell } \mathring g_{ \ell}(r) +  {\mdanger \mathring{\zeta}}_{\ell}
             {\mathringB}_0 \mathring g_0(r)
 \right) \,.
\end{equation}

Letting
\begin{equation}\label{19I18.2E}
  E^z(r,\theta,z) = e^{i m \theta} \sum_{\ell \in \Z}
    \smalle^z_\ell (r) e^{i( \ell  \beta_1 {-\beta} )z }
  \,,
\end{equation}
an identical calculation gives
\begin{equation}\label{20I18.+3+}
 \mbox{for $\ell\in \{0,\pm 1\}$, }
 \
 \smalle^z_\ell(r) =
 \varepsilon^{|\ell|}
 \left(
         {\mathringA}_{\ell } \mathring g_{ \ell}(r) +  {\mdanger \mathring{\zeta}}_{\ell}
             {\mathringA}_0 \mathring g_0(r)
 \right) \,.
\end{equation}
with constants $ {\mathringA}_\ell$
satisfying
\begin{equation}\label{9III18.11+}
  | {\mathringA}_{ \pm 1}|   \varepsilon
\le  {\varepsilon_1} | {\mathringA}_{0}|
 \,.
\end{equation}
We return to \eqref{pot:Fara}-\eqref{pot:Max}, which  for time-periodic fields take the form
\begin{align}
i \omega \psi(z)\mu_0 \vec{H}(\vec{r},t)&=-\nabla \times \vec{E}(\vec{r},t) -\frac{2}{c^2}\left(\nabla\phi \times \vec{E}(\vec{r},t)\right) \label{pot:Fara+}
 \,,
\\
i \omega \psi(z)\epsilon_0n^2(r) \vec{E}(\vec{r},t)&=\nabla \times \vec{H}(\vec{r},t) +\frac{2}{c^2}\left(\nabla\phi \times \vec{H}(\vec{r},t)\right) \label{pot:Max+}
 \,.
\end{align}
Using Cartesian coordinates, this can be rewritten as
\begin{align}
 H^x=& -\frac{i}{\psi k  }\sqrt{\frac{\epsilon_0}{\mu_0}}
  \left(\partial_z E^y- \partial_{y}E^z
   +  { \frac{2 \partial_z \phi}{  c^2} } E^y\right)
   \,,
\\
 H^y=&
 -\frac{i}{\psi k  } \sqrt{\frac{\epsilon_0}{\mu_0}}
  \left( \partial_xE^z-  \partial_z E^x
   -  { \frac{2 \partial_z \phi}{  c^2} } E^x\right)
   \,,
\\
 H^z=&
 -\frac{i}{\psi k }\sqrt{\frac{\epsilon_0}{\mu_0}}
  \left( \partial_{y}E^x {-}\partial_x E^y\right)
   \,,
\\
 E^x=&
   \frac{i}{\psi k n^2}\sqrt{\frac{\mu_0}{\epsilon_0}}
  \left(
   \partial_z H^y- \partial_{y}H^z
   +  { \frac{2 \partial_z \phi}{ c^2}  } H^y
    \right)
   \,,
\\
 E^y=&
 \frac{i}{\psi k n^2} \sqrt{\frac{\mu_0}{\epsilon_0}}
  \left( \partial_xH^z-  \partial_z H^x
  -  { \frac{2 \partial_z \phi}{ c^2}  } H^x
    \right)
   \,,
    \label{24VI18.23}
\\
 E^z=&
 \frac{i}{\psi k n^2}\sqrt{\frac{\mu_0}{\epsilon_0}}
  \left(\partial_{y}H^x {-}\partial_x H^y\right)
   \,.
\end{align}
Proceeding as in Section~\ref{ss1II18.1}, we are led to%
\footnote{One could streamlime somewhat the equations that follow by getting rid of the field $E^y$ in the right-hand side of \eqref{29I18.4+} using \eqref{24VI18.23}, similarly for the remaining equations. This gives more elegant but somewhat longer equations.}
\begin{align}
 H^x
  +
      &
   \frac{1}{\psi k} \partial_z\left(\frac{\partial_z H^x} { \psi k n^2}
    \right)
       +
       \frac{2 }{\psi k  c^2 }
    \partial_z\left( \frac{  \partial_z \phi  H^x}{\psi k n^2} \right)
   \nn
\\
 &=
     \frac{i}{\psi k}\sqrt{\frac{\epsilon_0}{\mu_0}}
      \left(
       \partial_{y}E^z
   -  \frac{2 \partial_z \phi}{ c^2}   E^y
    \right)
  + \frac 1 {\psi k}  \partial_z\left(\frac{\partial_x H^z} { \psi k n^2}
    \right)
 \,,
 \label{29I18.4+}
\\
 H^y
  +
      &
   \frac{1}{\psi k} \partial_z\left(\frac{\partial_z H^y} { \psi k n^2}
    \right)
      +
      \frac{2 }{\psi k  c^2 }
    \partial_z\left( \frac{  \partial_z \phi  H^y}{\psi k n^2} \right)
   \nn
\\
 &=
    -
     \frac{i}{\psi k}\sqrt{\frac{\epsilon_0}{\mu_0}}
      \left(
       \partial_{x}E^z
   -  \frac{2 \partial_z \phi}{ c^2}   E^x
    \right)
  + \frac 1 {\psi k}  \partial_z\left(\frac{\partial_y H^z} { \psi k n^2}
    \right)
 \,,
\\
E^x
  +
    &
   \frac{1}{\psi k {n^2}} \partial_z\left(\frac{\partial_z E^x} { \psi k}
    \right)
   +
   \frac{2 }{\psi k   {n^2} c^2 }
    \partial_z\left( \frac{  \partial_z \phi  E^x}{\psi k} \right)
   \nn
\\
   &
     =
         -\frac{i}{\psi k n^2 }\sqrt{\frac{\mu_0}{\epsilon_0}}
        \left(
  \partial_{y}H^z -   \frac{2 \partial_z \phi}{  c^2}  H^y
  \right)
  +
   \frac 1 {\psi k n^2} \partial_z \left( \frac{\partial_x E^z}{\psi k}
    \right)
 \,,
\\
 E^y
  +
      &
   \frac{1}{\psi k n^2} \partial_z\left(\frac{\partial_z E^y} { \psi k}
    \right)
      +
      \frac{2 }{\psi k  c^2 n^2 }
    \partial_z\left( \frac{  \partial_z \phi  E^y}{\psi k} \right)
   \nn
\\
 &=
     \frac{i}{\psi k n^2}\sqrt{\frac{\mu_0}{\epsilon_0}}
      \left(
       \partial_{x}H^z
   -  \frac{2 \partial_z \phi}{ c^2}   H^x
    \right)
  + \frac 1 {\psi k n^2}  \partial_z\left(\frac{\partial_y E^z} { \psi k}
    \right)
\,.
 \label{29I18.5+}
\end{align}

Similarly to \eqref{19I18.2} and to \eqref{19I18.2E}, we choose $m\in \Z$ and
we  seek a solution of the form
\begin{eqnarray}
  \vec H(r,\theta,z) &= &  \sum_{\ell \in \Z}
    \vec{\smallh}_\ell (r,\theta) e^{i( \ell  \beta_1{-\beta} )z }
\equiv
    e^{i m \theta} \sum_{\ell \in \Z}
    \vec{\smallh}_\ell (r) e^{i( \ell  \beta_1{-\beta} )z }
    \,,\label{19I18.2+B}
\\
  \vec E(r,\theta,z)
   &  = & \sum_{\ell \in \Z}
    \vec{\smalle}_\ell (r,\theta) e^{i( \ell  \beta_1{-\beta} )z }
\equiv
    e^{i m \theta} \sum_{\ell \in \Z}
    \vec{\smalle}_\ell (r) e^{i( \ell  \beta_1{-\beta} )z }
  \,.
   \label{19I18.2+E}
\end{eqnarray}

As before, we will consider solutions such that
\begin{equation}\label{3II18.1}
  \smallh^i_\ell = O(\varepsilon_1^{\min(|\ell|,2)} |\mathringB_0|)
  \,,
  \quad
  \smalle^i_\ell = O(\varepsilon_1^{\min(|\ell|,2)} |\mathringA_0|)
  \,.
\end{equation}

In all calculations that follow, $\smallh^i_\ell$ and  $\smalle^i_\ell$ will stand for $\smallh^i_\ell(r,\theta)$  and  $\smalle^i_\ell(r,\theta)$.

Inserting \eqref{19I18.2+B}-\eqref{3II18.1} into \eqref{29I18.4+}-\eqref{29I18.5+} one immediately finds that the equations for the fields $\mathring\smallh^i_0$ and $\mathring\smalle^i_0$ can be handled in a
manner identical to that of Section~\ref{ss1II18.2}, resulting in  $\beta$ being determined by the value of the gravitational potential on the axis of the spool.
In particular all the formulae derived there for the dependence of $\beta$ upon the height $h$ of the axis of the spool relative to a reference location apply.
Moreover, the calculations there suggest strongly that we can safely ignore the periodic variation of the coordinate radius of the core along the waveguide, and thus assume that the physical radius coincides with the coordinate radius  $a$  and is constant throughout the waveguide.

With our values of parameters, when $\aone=1$ we find
%
\begin{equation}\label{8III18.1}
  \mathringB_0 = -0.0038891 \, i \, \mathringA_0
   \,.
\end{equation}
%

To determine the remaining components of the fields,
it is convenient to multiply all equations \eqref{29I18.4+}-\eqref{29I18.5+} by $\psi^2 k^2 n^2$. After this multiplication, the operator appearing at the left-hand side of \eqref{29I18.4+} becomes, neglecting as usual $O(\epsilon^2)$ or higher order terms and second derivatives of $\phi$,
%
\begin{eqnarray}
 \nonumber
 \lefteqn{
 \psi^2 k^2 n^2  H^x
  +
    \psi \partial_z\left(\frac{\partial_z H^x} { \psi }
    \right)
       +
      \frac{2 \psi}{ c^2 }
    \partial_z\left( \frac{  \partial_z \phi  H^x}{\psi } \right)
   }
   &&
\\
 &\approx&
    \left(1- \frac{2 \phi }{c^2}\right)^2 k^2 n^2 H^x
  +
     \partial_z^2 H^x
  +
     \frac{4 \partial_z \phi }{c^2 }
     \partial_z H^x
\nn
\\
    & {\approx}&
     \partial_z^2 H^x
   {-}
     4  \varepsilon
   {\noatwo} \sin(\beta_1 z )
     \partial_z H^x
  +
    \big({\aonesquared }\plustwoatwo \varepsilon  \cos(\beta_1 z )\big)
     k^2 n^2 H^x
     \nonumber
\\
  &=&
   \sum_{\ell \in \Z}
   \Big[
   \left(
  \aonesquared  k^2 n^2-\left(\beta-\ell\beta_1\right)^2\right) \smallh^x_\ell
    {+}{
   2\beta_1\varepsilon( \ell  \beta_1{-\beta} )\danger
    \left(\smallh^x_{\ell+1}
     - \smallh^x_{\ell-1}\right)}
          \nonumber
\\
 &&
 +
     2\varepsilon \beta_1^{ {2}}
    (  \smallh^x_{\ell-1} + \smallh^x_{\ell+1})
 -
   2\varepsilon n^2 k^2 \left(\smallh^x_{\ell-1}+\smallh^x_{\ell+1}\right)
      \Big] e^{i ( \ell  \beta_1 {-\beta} )  z}
     \,.
    \label{3II18.2}
\end{eqnarray}
At the same level of accuracy, the right-hand side of \eqref{29I18.4+} becomes
%
\begin{eqnarray}
 \nonumber 
  \lefteqn{  {i}{\psi k n^2 }\sqrt{\frac{\epsilon_0}{\mu_0}}
      \left(
       \partial_{y}E^z
   -  \frac{2 \partial_z \phi}{ c^2}   E^y
    \right)
  +  \psi \partial_z\left(\frac{\partial_x H^z}{\psi}
    \right)
    }
     &&
     \\
  &\approx&  {i}{ k n^2 }\sqrt{\frac{\epsilon_0}{\mu_0}}
      \left(
       {\left({\aone}-2\varepsilon\cos(\beta_1z)\right)} \partial_{y}E^z
   -  \frac{2 \partial_z \phi}{ c^2}   E^y
    \right)
  +  \psi \partial_z\left(\frac{\partial_x H^z}{\psi}
    \right)
     \nonumber
\\
  &\approx & \partial_z\partial_x H^z
    + {i}{ k n^2 }\sqrt{\frac{\epsilon_0}{\mu_0}}
       {\left({\aone}-2\varepsilon\cos(\beta_1z)\right)} \partial_{y}E^z
          \nonumber
\\
&&
     {-} 2 \varepsilon \beta_1 \sin(\beta_1 z)
   \left(\partial_x H^z
      -{i}{ k n^2 }\sqrt{\frac{\epsilon_0}{\mu_0}}
      E^y
    \right)
     \nonumber
\\
  &=&
   i
   \sum_{\ell \in \Z}
   \Big[
   ( \ell  \beta_1{-\beta} )\danger \partial_x \smallh^z_\ell
    + { k n^2 }\sqrt{\frac{\epsilon_0}{\mu_0}}
      {\aone} \partial_{y}\smalle^z_\ell
      \nonumber
\\
 &&
    +
     \varepsilon \beta_1
    \big( \partial_x \smallh^z_{\ell-1} - \partial_x \smallh^z_{\ell+1}
      -{i}{ k n^2 }\sqrt{\frac{\epsilon_0}{\mu_0}}
      (\smalle^y_{\ell-1} - \smalle^y_{\ell+1})
      \big)
            \nonumber
\\
 &&
  {-\varepsilon k n^2 \sqrt{\frac{\epsilon_0}{\mu_0}}\left(\partial_y\smalle^z_{\ell-1}+\partial_y\smalle^z_{\ell+1}\right)}
      \Big] e^{i ( \ell  \beta_1 {-\beta} )  z}
     \,.
    \label{7II18.1}
\end{eqnarray}
Hence
\begin{align}
&\left(\aonesquared  k^2 n^2- \left(\beta-\ell\beta_1\right)^2\right) \smallh^x_\ell
  {+}{2\beta_1\varepsilon( \ell  \beta_1{-\beta} )\danger
  \left(\smallh^x_{\ell+1}
  -
  \smallh^x_{\ell-1}\right)}
+2\varepsilon \beta_1^{ {2}}(\smallh^x_{\ell-1}
 +
 \smallh^x_{\ell+1})
 \nonumber
\\
  &-2\varepsilon n^2 k^2 \left(\smallh^x_{\ell-1}+\smallh^x_{\ell+1}\right)
  =
   { i {\nodanger} \Big[
  }
 ( \ell  \beta_1{-\beta} )\danger \partial_x \smallh^z_\ell+{ k n^2 }\sqrt{\frac{\epsilon_0}{\mu_0}}
      {\aone} \partial_{y}\smalle^z_\ell
      \nonumber
\\
 &+\varepsilon \beta_1\big( \partial_x \smallh^z_{\ell-1} -\partial_x \smallh^z_{\ell+1}-{i}{ k n^2 }\sqrt{\frac{\epsilon_0}{\mu_0}}(\smalle^y_{\ell-1} - \smalle^y_{\ell+1})\big)
  {
  -\varepsilon k n^2 \sqrt{\frac{\epsilon_0}{\mu_0}}
   \left(\partial_y\smalle^z_{\ell-1}
   +
   \partial_y\smalle^z_{\ell+1}\right)}
   \Big]
    \,.
\label{8II18.1}
\end{align}
In particular, ignoring terms of order $O(\epsilon_1^2)$,
%
\begin{eqnarray}
\lefteqn{
   \left(
    \aonesquared  k^2 n^2
   -
    {(\beta_1-\beta)^2}
   \right)
    \smallh^x_1
   =
  i{\nodanger}
   (-\beta+ \beta_1) \partial_x \smallh^z_1
    +
  i{\nodanger}
  { k n^2 }\sqrt{\frac{\epsilon_0}{\mu_0}}
      {\aone} \partial_{y}\smalle^z_1
      \nonumber
      }
      &&
\\
     &&
     +
     \varepsilon
     \Big(2(    k^2 n^2 -\beta_1^{ {2}}
       {+}{
       \beta_1\left(-\beta+\beta_1\right)}
    )
      \smallh^x_{0}
    +
  i{\nodanger}
  \beta_1
    \big( \partial_x \smallh^z_{0}
      -{i}{ k n^2 }\sqrt{\frac{\epsilon_0}{\mu_0}}
      \smalle^y_{0}
      \big) {-
  i{\nodanger}
  kn^2\sqrt{\frac{\epsilon_0}{\mu_0}}\partial_y\smalle^z_{0}}
      \Big)
     \,,
     \nn
\\
 &&
\\
\lefteqn{
   \left(
    \aonesquared  k^2 n^2
   -
    {(\beta_1+\beta)^2}
   \right)
    \smallh^x_{-1}
    =
  i{\nodanger}
   (-\beta- \beta_1) \partial_x \smallh^z_{-1}
    +
  i{\nodanger}
  { k n^2 }\sqrt{\frac{\epsilon_0}{\mu_0}}
      {\aone} \partial_{y}\smalle^z_{-1}
      \nonumber
      }
      &&
\\
     &&
     +
     \varepsilon
     \Big(2(    k^2 n^2 -\beta_1^{ {2}}
      {+}{
       \beta_1\left(\beta+\beta_1\right)}
    )
      \smallh^x_{0}
   -
  i{\nodanger}
  \beta_1
    \big(  \partial_x \smallh^z_{0}
      -{i}{ k n^2 }\sqrt{\frac{\epsilon_0}{\mu_0}}
      \smalle^y_{0}
      \big) {-
  i{\nodanger}
  kn^2\sqrt{\frac{\epsilon_0}{\mu_0}}\partial_y\smalle^z_{0}}
      \Big)
     \,,
     \nn
\\
&&
    \label{8II18.2}
\end{eqnarray}
which determines $\smallh^x_{\pm 1}$.
Similarly one finds,  {for $\ell=\pm 1$,}
\begin{eqnarray}
\lefteqn{
   \left(
    \aonesquared  k^2 n^2
   -
    {(\beta-\ell\beta_1 )^2}
   \right)
    \smallh^y_{\ell}
    =
  i{\nodanger}
   (-\beta+ \ell \beta_1) \partial_y \smallh^z_{\ell}
    -
  i{\nodanger}
   { k n^2 }\sqrt{\frac{\epsilon_0}{\mu_0}}
       {\aone}
       \partial_{x}\smalle^z_{\ell}
      \nonumber
      }
      &&
\\
     &&
     +
     \varepsilon
     \Big(2(    k^2 n^2
     -\beta_1^{ {2}}  - {\beta_1\left({ {\ell}}\beta-\beta_1\right)}
    )
      \smallh^y_{0}
    + \ell
  i{\nodanger}
  \beta_1
    \big( \partial_y \smallh^z_{0}
      +{i}{ k n^2 }\sqrt{\frac{\epsilon_0}{\mu_0}}
      \smalle^x_{0}
      \big)
       {
      +
  i{\nodanger}
   kn^2\sqrt{\frac{\epsilon_0}{\mu_0}}\partial_x\smalle^z_{0}}
      \Big)
     \,.
     \nn
\\
 &&
    \label{9II18.1}
\end{eqnarray}
Equivalently, in an orthonormal frame adapted to cylindrical symmetry (compare \eqref{++equ:Er+}-\eqref{++equ:Etheta+}),
\begin{eqnarray}
\lefteqn{
   \left(
    \aonesquared  k^2 n^2
   -
   {(\beta -{ {\ell}}\beta_1)^2}
   \right)
    \smallh^r_{ {\ell}}
    =
  i{\nodanger}
   (-\beta+ { {\ell}}\beta_1)
   \partial_r \smallh^z_{ {\ell}}
    +
  i{\nodanger}{ k n^2 }\sqrt{\frac{\epsilon_0}{\mu_0}}
      {\aone}
      \frac 1 r \partial_{\theta}\smalle^z_{ {\ell}}
      \nonumber
      }
      &&
\\
     &&
     +
     \varepsilon
     \Big(2(    k^2 n^2 -\beta_1^{ {2}}
       -
     {\beta_1\left({ {\ell}}\beta-\beta_1\right)}
    )
      \smallh^r_{0}
    +{ {\ell}}
  i{\nodanger}
  \beta_1
    \big( \partial_r \smallh^z_{0}
       {-}{i}{ k n^2 }\sqrt{\frac{\epsilon_0}{\mu_0}}
      \smalle^\theta_{0}
      \big) {
      -
  { i }
      kn^2\sqrt{\frac{\epsilon_0}{\mu_0}}\frac{1}{r}\partial_{\theta}\smalle^z_{0}}
      \Big)
     \,,
 \nn
\\
 &&
    \label{9II18.2}
\\
 \lefteqn{
   \left(
    \aonesquared  k^2 n^2
   -
   {(\beta-{ {\ell}}\beta_1)^2}
   \right)
    \smallh^\theta_{ {\ell}}
    =
  i{\nodanger}
   (-\beta+ { {\ell}}\beta_1)
   \frac 1 r \partial_\theta \smallh^z_{ {\ell}}
    -
  i{\nodanger}
  { k n^2 }\sqrt{\frac{\epsilon_0}{\mu_0}}
      {\aone}
      \partial_{r}\smalle^z_{ {\ell}}
      \nonumber
      }
      &&
\\
     &&
     +
     \varepsilon
     \Big(2(    k^2 n^2 -\beta_1^{ {2}}
       -
     {\beta_1\left({ {\ell}}\beta-\beta_1\right)}
    )
      \smallh^\theta_{0}
    +{ {\ell}}
  i{\nodanger}
  \beta_1
    \big(
     \frac 1 r \partial_\theta \smallh^z_{0}
       {+}{i}{ k n^2 }\sqrt{\frac{\epsilon_0}{\mu_0}}
      \smalle^r_{0}
      \big) {+
  i{\nodanger}
  kn^2\sqrt{\frac{\epsilon_0}{\mu_0}}\partial_r\smalle^z_{0}}
      \Big)
     \,.
   \nn
\\
 &&
      \label{9II18.4}
\end{eqnarray}

Following the same steps  for the electric field components one finds, again for $\ell=\pm 1$,
%
\begin{eqnarray}
\lefteqn{
   \left(
    \aonesquared  k^2 n^2
   -
   {(\beta-{ {\ell}}\beta_1)^2}
   \right)
    \smalle^r_{ {\ell}}
    =
  i{\nodanger}
   (-\beta+ { {\ell}}\beta_1)
   \partial_r \smalle^z_{ {\ell}}
    -  { i } {k}
  \sqrt{\frac{\mu_0}{\epsilon_0}}
      {\aone}
      \frac 1 r \partial_{\theta}\smallh^z_{ {\ell}}
      \nonumber
      }
      &&
\\
     &&
     +
     \varepsilon
     \Big(2(    k^2 n^2 -\beta_1^{ {2}}
       -
     {\beta_1\left({ {\ell}}\beta-\beta_1\right)}
    )
      \smalle^r_{0}
    +{ {\ell}}
  { i }
  \beta_1
    \big( \partial_r \smalle^z_{0}
       {+}{i}{k}\sqrt{\frac{\mu_0}{\epsilon_0}}
      \smallh^\theta_{0}
      \big)
       {+
   { i }
  k\sqrt{\frac{\mu_0}{\epsilon_0}}\frac{1}{r}\partial_{\theta}\smallh^z_{0}}
      \Big)
     \,,
 \nn
\\
 &&
    \label{9II18.+2}
\\
 \lefteqn{
   \left(
    \aonesquared  k^2 n^2
   -
   {(\beta-{ {\ell}}\beta_1)^2}
   \right)
    \smalle^\theta_{ {\ell}}
    =
  i{\nodanger}
   (-\beta+ { {\ell}}\beta_1)
   \frac 1 r \partial_\theta \smalle^z_{ {\ell}}
    +
  i{\nodanger}
  {k}\sqrt{\frac{\mu_0}{\epsilon_0}}
      {\aone}
      \partial_{r}\smallh^z_{ {\ell}}
      \nonumber
      }
      &&
\\
     &&
     +
     \varepsilon
     \Big(2(    k^2 n^2 -\beta_1^{ {2}}
       -
     {\beta_1\left({ {\ell}}\beta-\beta_1\right)}
    )
      \smalle^\theta_{0}
    +{ {\ell}}
  i{\nodanger}
  \beta_1
    \big(
     \frac 1 r \partial_\theta \smalle^z_{0}
       {-}{i}{ k}\sqrt{\frac{\mu_0}{\epsilon_0}}
      \smallh^r_{0}
      \big)
       {-
  i{\nodanger}
  k\sqrt{\frac{\mu_0}{\epsilon_0}}\partial_r\smallh^z_{0}}
      \Big)
     \,.
   \nn
\\
 &&
      \label{9II18.+4}
\end{eqnarray}
For any choice of $ {\mathringA}_{ {\ell}}$ and $ {\mathringB}_{ {\ell}}$ we can use \eqref{9II18.2}-\eqref{9II18.+4} to determine $\smallh_{ {\ell}}^r$, $\smallh_{ {\ell}}^\theta$, $\smalle_{ {\ell}}^r$ and $\smalle_{ {\ell}}^\theta$. However, the resulting fields will not be continuous in general. We thus need to make sure that there is a choice $( {\mathringA}_{ {\ell}}, {\mathringB}_{ {\ell}})$ which renders the relevant fields continuous, as necessary for a solution of all equations.
The vanishing of surface currents requires continuity of $H^\theta$ and $E^\theta$.
For this, let us write \eqref{9II18.4} and \eqref{9II18.+4} as
\begin{equation}\label{10II18.1}
  \left(
  \begin{array}{c}
    \smalle^{\theta}_{ {\ell}} \\
   \smallh^\theta_{ {\ell}}
 \end{array}
 \right)
    = L( {\mathringA}_{ {\ell}}, {\mathringB}_{ {\ell}}) + s_{ {\ell}}
    \,,
\end{equation}
where $L$ is the part of the solution which involves $( {\mathringA}_\ell, {\mathringB}_\ell)$.
  For any function $f$ we set
\begin{equation}\label{10II18.2}
  f_+:=\lim_{r\searrow a} f(r)
  \,,
  \quad
  f_-:=\lim_{r\nearrow a} f(r)
  \,,
  \quad
  [f] = f_+ - f_-
  \,.
\end{equation}

In subsequent calculations the following relations are useful:
\begin{align}
 &
 \frac{(\beta-\ell \beta_1)^2}{k^2}= \aonesquared  n_1^2-\frac{U_\ell^2}{a^2k^2}
  \,,
\\
 &
 \frac{(\beta-\ell \beta_1)^2}{k^2}= \aonesquared  n_2^2+\frac{W_\ell^2}{a^2k^2}
  \,
\\
 &
 \frac{(\beta-\ell \beta_1)^2}{\aonesquared  k^2}\left(\frac{1}{U_\ell^2}
  +\frac{1}{W_\ell ^2}\right)=\left(\frac{n_1^2}{U_\ell^2}+\frac{n_2^2}{W_\ell^2}\right)
 \,.
\end{align}

The requirement of continuity of $\smalle^{\theta}_{ {\ell}}$ and $\smallh_{ {\ell}}^\theta$ leads to the system of equations
%
\begin{equation}
  {a \varepsilon} i{\nodanger}
 \left(
 \begin{matrix}
 im(\beta -\ell \beta_1) \left(\frac{1}{U^2_{ {\ell}}}+\frac{1}{W^2_{ {\ell}}}\right)
   &
  -\sqrt{\frac{\mu_0}{\epsilon_0}} {{\aone} k }  \left(\frac{1}{U_{ {\ell}}}\frac{J'_m(U_{ {\ell}})}{J_m(U_{ {\ell}})}
  +\frac{1}{W_{ {\ell}}}\frac{ K'_m(W_{ {\ell}}) }{ K_m(W_{ {\ell}}) }\right)
\\
 \sqrt{\frac{\epsilon_0}{\mu_0}} {\aone}  k \left(\frac{n_1^2}{U_{ {\ell}}}\frac{J'_m(U_{ {\ell}})}{J_m(U_{ {\ell}})}
  +\frac{n_2^2}{W_{ {\ell}}}\frac{ K'_m(W_{ {\ell}}) }{ K_m(W_{ {\ell}}) }\right)
   &
    i m (\beta-\ell\beta_1) \left(\frac{1}{U_{ {\ell}}^2}+\frac{1}{W_{ {\ell}}^2}\right)
\end{matrix}
\right)
 \left(
 \begin{matrix}
 {\mathringA}_{ {\ell}}\\
 {\mathringB}_{ {\ell}}
\end{matrix}
\right)
+[s_{ {\ell}}] = 0
 \,.
  \label{7III18.1asdf}
\end{equation}
A unique solution will exist for any jump $[s]$ of the source term $s$ provided the matrix above has a non-vanishing determinant. The latter reads, with $\ell\in \{0,\pm 1\}$:
\begin{eqnarray}
 \nn
 \lefteqn{
F_\ell (\beta-\ell\beta_1):=  \Bigg\{{m^2(\beta -\ell \beta_1)^2}\left(\frac{1}{U_\ell ^2}+\frac{1}{W_\ell^2}\right)^2
}
&&
\\
&&
 -  \aonesquared  {  k^2}
 \left(\frac{1}{U_\ell}\frac{J'_m(U_\ell)}{J_m(U_\ell)}
 +\frac{1}{W_\ell}\frac{ K'_m(W_\ell) }{ K_m(W_\ell) }\right)
  \left(\frac{n_1^2}{U_\ell}\frac{J'_m(U_\ell)}{J_m(U_\ell)}+\frac{n_2^2}{W_\ell}\frac{ K'_m(W_\ell) }{ K_m(W_\ell) }\right)
  \Bigg\}
  \,.
\label{equ:ModeEquNOT}
\end{eqnarray}
Recall that $F_0(\beta)=0$ by \eqref{equ:ModeEqu}, and that $\beta_1$ is small compared to $\beta$. A {\sc Mathematica} plot$^{\textrm{\ref{foot2III18.1}}}$ presented in Figure~\ref{F2III18.1} shows that $F_\ell(\beta-\ell \beta_1)$, with $m=1$, $\ell=\pm 1$, and with parameter values \eqref{1XI17.11}-\eqref{31I18.1} and \eqref{19I18.3+}, does not vanish. We conclude that \eqref{7III18.1asdf} with $\ell\in\{\pm 1\}$ can indeed be solved.
\begin{figure}[ht]
	\centering
  \includegraphics[width=0.5\textwidth]{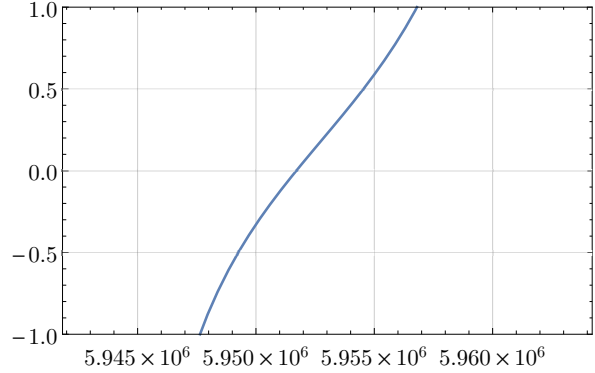}
	\caption{The determinant function $F_{-1} (\beta+ \beta_1)$, plotted as a function of $\beta+ \beta_1$, has an isolated zero at $\beta+\beta_1=\beta$. Note that $F_{-1} (\beta+ \beta_1)=F_{1} (\beta- \beta_1)$, so that the plot also proves that $F_{1}(\beta- \beta_1)$ vanishes only at $\beta_1=0$ in the interval of interest. }
	\label{F2III18.1}
\end{figure}

Consistency of our scheme requires that the constants $(\mathringA_{\pm 1}, \mathringB_{\pm 1})$ calculated above are of sufficiently controlled order. We have implemented the above calculation, with $m=1$ and the values of parameters given above, in {\sc Mathematica}, leading to
\begin{align}
  \varepsilon \mathringA_1
 &\approx
 -1.74791
\times 10^{-13}
 \mathringA_0
  \,, \nonumber
\\
\nn
 \varepsilon  \mathringB_1
 &\approx
 6.79388 \times 10^{-16} i \,
\mathringA_0
\,,
\\
\nn
 \varepsilon  \mathringA_{-1}
&\approx 1.25671 \times 10^{-13}
 \mathringA_0
  \,,
\\
 \varepsilon  \mathringB_{-1}
&\approx -4.88466 \times 10^{-16} i
\mathringA_0
\,,
\label{14III18.1}
\end{align}
where \eqref{8III18.1} has also been used.
We note that for $d_s\in[10^{-1}\mathrm{m},10\,\mathrm{m}]$ the fields scale approximately linearly with $d_s$. Thus, changing $d_s$ within this range has the effect of multiplying the above by $10 d_s/ \mathrm{m}$.

Summarising, we have obtained, after truncating to two significant digits,
%
\begin{eqnarray}
 \nonumber 
  E^z &\approx&  \mathringA_0  \times e^{i(\omega t + \theta - \beta z)}
  \Big[
  \mathring g_0(r)
\\
 \nonumber
 &&
    +\left(6.64  \,  \mathring g_0(r) - 0.17 \, \mathring g_1(r) \right) 10^{-12} \times e^{i\beta_1 z}
\\
 &&
  -
  \left(6.64 \,  \mathring g_0(r)  - 0.13  \,  \mathring g_{-1}(r)   \right) 10^{-12}\times e^{-i\beta_1 z}
  \Big]
  \,,
\\
\nn
 H^z
  &\approx& - 3.9 \times 10^{-3}\, i \,  \times  \mathringA_0 \times  e^{i(\omega t + \theta - \beta z)}
  \Big[
  \mathring g_0(r)
\\
 \nonumber
 &&
    - \left(6.64 \,  \mathring g_0(r) + 0.17  \,  \mathring g_1(r) \right) 10^{-12} \times e^{i\beta_1 z}
\\
 &&
  +
  \left(   6.64 \,  \mathring g_0(r) + 0.13 \, \mathring g_{-1}(r)\right) 10^{-12} \times e^{-i\beta_1 z}
  \Big]
  \,.
\end{eqnarray}
with $\beta$ given by \eqref{31I18.1}, $\beta_1 = 1/d_s$, where $d_s$ is the radius of the spool, $\mathring g_\ell $ given by \eqref{19I18.9}, and $\mathringA_0$ is a free constant determined by the amplitude of the entrant laser beam.

We see that the amplitude of the new modes created by the interaction with the gravitational potential is of the order of
$10^{-11}$ times the amplitude of the main $m=1$ modes for a spool with a radius of 30 centimeters.
The possibility of measuring  these additional modes
in a laboratory on earth using current technology is unclear,
given the power constraints in optical waveguides and the small difference in propagation constants.

We have checked that \eqref{14III18.1} guarantees continuity of $B^r$  up to four significant digits. One can likewise calculate explicitly the remaining fields, but the formulae are not very enlightening. In order to get an idea of the order of the solutions we calculate their values at the center of the waveguide, i.e.\ $r=0$. One finds:
\begin{eqnarray}
  E^r &
  \approx
   &
   \mathringA_0 \times  e^{i(\omega t + \theta - \beta z)}
  i
   ( - 27.78   +   9.06 \times 10^{-11} e^{-i \beta_1 z} -
     8.99\times 10^{-11} e^{i \beta_1 z})
   \,,
    \phantom{xxx}
\\
  E^\theta  &
  \approx
  &
   \mathringA_0 \times  e^{i(\omega t + \theta - \beta z)}
 (27.78 - 9.04 \times10^{-11} e^{-i \beta_1 z} + 8.97\times10^{-11} e^{i \beta_1 z})
   \,,
\\
  H^r &
  \approx
   &
   \mathringA_0 \times  e^{i(\omega t + \theta - \beta z)}
 (
   -0.11 + 3.59   \times 10^{-13} e^{- i \beta_1  z} -
   3.50 \times 10^{-13} e^{  i \beta_1 z}
   )
   \,,
\\
  H^\theta  &
  \approx
  &
   \mathringA_0 \times  e^{i(\omega t + \theta - \beta z)} i
 (-0.11  +   3.61 \times 10^{-13}  e^{-i \beta_1 z} -
     3.61 \times 10^{-13}  e^{i \beta_1 z})
   \,.
\end{eqnarray}
%






\appendix

\section{Static metrics}
 \label{A11II18.1}

In this appendix we provide a geometric decomposition of the Maxwell equations in terms of electric and magnetic fields for general static metrics.

Let $u^\alpha$ be a  unit vector collinear with a timelike static Killing vector $\xi^\alpha$,
 thus it holds that
 \begin{equation}\label{14VII17.1}
   \xi_{[\alpha}\nabla_\beta\xi_{\gamma]}
    = 0
    \,.
 \end{equation}
A standard calculation based on this equation shows that
\begin{equation}\label{static}
\nabla_\alpha u_\beta = - u_\alpha a_\beta\,,
\end{equation}
with
$$
a_\alpha = \nabla_u u_\alpha = (- \xi^2)^{- \frac{1}{2}} {\Driem}_\alpha (- \xi^2)^\frac{1}{2} = {\Driem}_\alpha \log((- \xi^2)^\frac{1}{2})
 \,,
$$
where ${\Driem}_\alpha$ is the covariant derivative operator of the metric in the horizontal space,
  i.e.\ the space orthogonal to $u$.

The Maxwell field can be uniquely written as the sum of an  associated electric and  magnetic contribution, as follows:
\begin{equation}
 \label{2XI17.6+}
F_{\alpha\beta} = 2 u_{[\alpha} \BobbyE_{\beta]} + \BobbyB _{\alpha\beta}\,,\hspace{0.6cm}\BobbyE_\alpha u^\alpha = 0
 \,,\,\,\,\,\,\,\BobbyB _{\alpha\beta} u^\alpha = 0
  \,,
\end{equation}
where
\begin{equation}\label{2XI17.7+}
 \fbox{$
 \BobbyE_\alpha = F_{\alpha\beta}u^\beta
 \,, \quad \BobbyB _{\alpha\beta} = - 3 F_{[\alpha\beta}u_{\gamma]} u^\gamma
\,.
$}
\end{equation}
The Maxwell equations read
\begin{equation}\label{homog+}
\nabla_{[\alpha}F_{\beta\gamma]} = \nabla_{[\alpha}(\BobbyB _{\beta\gamma]} + 2 u_\beta \BobbyE_{\gamma]}) = 0
 \,,
\end{equation}
and, assuming that both $\epsilon$ and $\mu$ are  point-independent,
\begin{equation}\label{station1+}
\nabla_\alpha [F^{\alpha\beta} - 2 (1- n^2)u^{[\alpha} F^{\beta]}{}_\rho u^\rho] = \nabla_\alpha \BobbyB ^{\alpha \beta} + 2 n^2 \nabla_\alpha(u^{[\alpha} \BobbyE^{\beta]}) = 0\,.
\end{equation}
Using \eqref{static}
 and the vanishing of the divergence of $u$, \eqref{station1+} can be written as
\begin{equation}\label{station2'}
\nabla_\alpha \BobbyB ^{\alpha\beta} + 2 n^2 u^{[\alpha}\nabla_\alpha \BobbyE^{\beta]} = 0
 \,.
\end{equation}
Taking the horizontal projection of (\ref{homog+}) (which means: contracting with $h^\alpha{}_\beta := \delta^\alpha{}_\beta + u^\alpha u_\beta$ on all indices) gives
\begin{equation}\label{magnetic'}
{\Driem}_{[\alpha} \BobbyB _{\beta\gamma]} = 0
 \,.
\end{equation}
Next, contracting (\ref{station2'}) with $u_\beta$ yields
\begin{equation}\label{electric'}
0= i_u \,d  (\BobbyB  + u \wedge \BobbyE) = \mathcal{L}_u \BobbyB  + a \wedge \BobbyE +  a \wedge \BobbyE +
 d\BobbyE
  + u \wedge (i_u d \BobbyE) \,.
\end{equation}
%
Equivalently, writing
\begin{equation}\label{2XII17.9}
 \dot \BobbyB  : = \mathcal{L}_\xi \BobbyB
 \,,
  \
  \mbox{thus}
 \
 \mathcal{L}_u \BobbyB _{\alpha\beta}=(- \xi^2)^{- \frac{1}{2}} \dot{\BobbyB }_{\alpha\beta}
  \,,
\end{equation}
where $\mathcal{L}$ denotes the Lie-derivative in space-time, we have
\begin{equation}
\dot{\BobbyB }_{\alpha\beta} =- 2 {\Driem}_{[\alpha}((- \xi^2)^\frac{1}{2} \BobbyE_{\beta]})\,.
\end{equation}
(We emphasise that the identity $\mathcal{L}_u \alpha = (- \xi^2)^{- \frac{1}{2}} \mathcal{L}_\xi \alpha$,
 used in \eqref{2XII17.9},  holds for horizontal differential forms, and that care must be taken when other tensor fields are considered.)

We finally take the horizontal projection of (\ref{station2'}). Using
\begin{equation}
\mathcal{L}_u \BobbyE_\alpha  \equiv  h_\alpha{}^\beta   \mathcal{L}_u \BobbyE_\beta = h_\alpha{}^\beta   \nabla_u \BobbyE_\beta
\end{equation}
and
\begin{equation}
h_\beta{}^\gamma \nabla^\alpha \BobbyB _{\alpha \gamma} = {\Driem}^\alpha \BobbyB _{\alpha\beta} + a^\alpha \BobbyB _{\alpha\beta} = (- \xi^2)^{- \frac{1}{2}} {\Driem}^\alpha ((- \xi^2)^\frac{1}{2}
\BobbyB _{\alpha\beta})
\end{equation}
we find that
\begin{equation}
n^2 \dot{\BobbyE}_\alpha = {\Driem}^\beta ((- \xi^2)^\frac{1}{2}
\BobbyB _{\alpha\beta})
 \,.
\end{equation}

Summarising, in the   notation of \eqref{2XI17.7+} the Maxwell constraint equations read
\begin{equation}\label{2XII17.8'}
  \fbox{${\Driem}_{[\alpha} \BobbyB _{\beta\gamma]} = 0
   \,,
   \quad  {\Driem}_\alpha \BobbyE^\alpha = 0\,,
  $
  }
\end{equation}
while the evolution equations take the form
\begin{equation}
\fbox{
 $
 \dot{\BobbyB }_{\alpha\beta} =- 2 {\Driem}_{[\alpha}((- \xi^2)^\frac{1}{2} \BobbyE_{\beta]})\,,
 \quad
  n^2 \dot{\BobbyE}_\alpha = {\Driem}^\beta ((- \xi^2)^\frac{1}{2}
 \BobbyB _{\alpha\beta})
 \,,
 $}
  \label{2XI17.10++}
\end{equation}
and recall that ${\Driem}$
 is the covariant derivative operator of the metric induced by the space-time metric on the leaves of the foliation orthogonal to the static Killing vector.

In a coordinate system in which $\xi^\alpha\partial_\alpha=\partial_t$ and $u_\beta dx^\beta$ is proportional to $dt$,  Equations~\eqref{2XI17.10++} coincide with \eqref{16VII17.14a}-\eqref{16VII17.14b} using the correspondence
\begin{equation}\label{27II18.21}
  \redE_k = c^{-1} \BobbyE_k
  \,,
  \quad
  \redB^i =  \frac 12 \epsilon^{ijk} \BobbyB_{jk}
  \,,
\end{equation}
with $\BobbyE_0=0=\BobbyB_{0\alpha}$.

\bigskip

\noindent{\sc Acknowledgements:} Special thanks are due to Stefan Palenta for detailed comments on a previous version of the manuscript. Useful discussions with  Peter Aichelburg
are acknowledged. RB and PTC are grateful to the Erwin Schr\"odinger Institute for hospitality  during part of work on this paper. CH is a recipient of a DOC Fellowship of the Austrian Academy of Sciences at the Institute of Physics. The research of PTC, CH and PW was  supported in
part by the Vienna University Research Platform TURIS.
PTC was further supported
by the Polish National Center of Science (NCN) under grant 2016/21/B/ST1/00940.

\providecommand{\bysame}{\leavevmode\hbox to3em{\hrulefill}\thinspace}
\providecommand{\MR}{\relax\ifhmode\unskip\space\fi MR }
\providecommand{\MRhref}[2]{%
  \href{http://www.ams.org/mathscinet-getitem?mr=#1}{#2}
}
\providecommand{\href}[2]{#2}

\end{document}